\documentclass[
 amsmath,amssymb,
]{revtex4-2}

\usepackage{graphicx}% Include figure files
\usepackage{orcidlink} %
\usepackage{xcolor} 
\usepackage{hyperref}
\hypersetup{
    colorlinks=true,      % false = boxed links; true = colored text
    linkcolor=blue,        % color of internal links (sections, equations)
    citecolor=blue,      % color of citations
    urlcolor=blue,     % color of external links / emails
}
\usepackage{physics}
\usepackage{dcolumn}% Align table columns on decimal point
\usepackage{bm}% bold math
\DeclareMathOperator{\sign}{sign}

\begin{document}

%\preprint{APS/123-QED}

\title{\textbf{Non-Markovian dynamics of giant  emitters beyond the Weisskopf–Wigner approximation} 
}% 
%\orcidlink{0000-0002-1734-1405}
\author{Carlos A. Gonz\'alez-Guti\'errez}
\email{carlosgg@icf.unam.mx}
\affiliation{Instituto de Ciencias Físicas, Universidad Nacional Autónoma de México, 62210, Mexico}

\date{\today}% It is always \today, today,

\begin{abstract}
Giant quantum emitters, whose effective size is comparable to the wavelength of the radiation they couple to, give rise to interference effects and non-Markovian dynamics that lie beyond the scope of the standard Weisskopf–Wigner approximation. Here we study minimal models of giant emitters coupled to cavity–array waveguides and 
identify symmetric configurations where the dynamics can be solved exactly. This allows us to capture the emergence of bound states both inside and outside the photonic continuum, and to demonstrate the possibility of engineering their coherent superpositions. Our results provide analytical insight into non-Markovian light–matter interaction and suggest a feasible implementation using superconducting circuit platforms.
\end{abstract}

\maketitle
%\tableofcontents

\section{Introduction}  
%Context
%Background\\
The theory of open quantum systems has been applied successfully to a broad variety of problems in quantum optics, solid state physics, quantum information and computation, and even to some dynamical processes in biological systems \cite{Weiss2012,BreuerBook2007,IdeVega2017}. This description often relies on the assumption of having a memoryless environment which allows the application of the so called Markovian approximation. A key quantity in this approximation is the system-environment spectral function that can be frequency-dependent in general. In the context of light-matter interaction in free space, this function is almost constant in the relevant range of optical frequencies, a fact that justifies the use of the Markovian approximation. The first theoretical approach to explain the process of spontaneous emission based on this assumption was provided by Weisskopf and Wigner in 1930 \cite{Weisskopf1930}. This formalism works perfectly fine and describes the irreversible exponential decay of atoms in most situations in free space. The situation is quite different when we deal with \emph{ structured} environments, showing non-trivial frequency behaviors of the spectral function \cite{GonzalezTudela&CiracPRA2017}. This is the case of quantum emitters coupling to photonic crystals \cite{Notomi_2010}, plasmonic waveguides \cite{GonzalezTudelaPorrasprl2013}, atomic waveguides \cite{AsenjoPRR2020}, transmission line superconducting circuits \cite{ScigliuzzoPRX2022,PainterPRX2021}, among others. The dynamics of the emitters turns non-Markovian and a treatment beyond the Weisskopf-Wigner approach is needed. As an example of the breakdown of the Markovian dynamics is the existence of bound states outside the continuum (BOC) when tuning the frequency of an emitter within a band gap \cite{SanchezBurilloPRA2017, GonzalezTudela&CiracPRA2017,CalajoPRA2016}. This leads to a partial (non-exponential) emission of the radiation into the medium, exponentially localizing the photonic wavefunction at the position of the emitter. In this case, the spectral function has two van Hove singularities at the band-edges, making the spectral function highly sensitive to frequency variations and therefore not amenable for a Markovian treatment. In order to have an appropriate description of a structured environment it becomes necessary to go beyond the standard Markovian treatment and to take into account the complexity of a realistic spectral function. This is indeed possible for some experimentally relevant  one-dimensional structured waveguides as we will see later in this paper.  
\\

In recent years artificial atoms known as \emph{giant atoms} have brought a lot of attention due to their interesting interference properties \cite{GuoPRA2017,KockumPRL2018,CarolloPRR2020,LonghiOPL2020,TaylorPRA2020,Kockumreview2021, WangKockumNoriPRL2021,FengPRA2021,DuPRA2021,HongweiPRA2021,KockumPRL2022,KockumPRA2022,Chen2022,Wang_2022,NoatcharPRA2022,TerradasPRA2022,Xiao2022,ChengPRA2022,GuoPRA2023,SoroKockumPRA2023,SantosPRL2023,YinPRA2023,PengPRA2023,HuatangPRA2023,Du_2023,Zheng_2023,ZhangPRA2023,Leong-Chuang2023,
RoccatiPRL2024,WangPRA2024,SoroPRR2024,NoriPRR224,LuoAQT2024,LiPRA2024,GaoxiangPRA2024,Xu_2024,HuaizhiPRA2024,GaoXiang2PRA2024,WengPRA2024,TaoPRA2024,XinPRA2024,Leonforte_2025,Chen_2025}. These artificial atoms are considered \emph{giant} since their effective size is comparable to the wavelength of the electromagnetic radiation they can couple to. As a consequence of this fact they can no longer be treated as point-like matter, and therefore the common dipole approximation becomes invalid. In the most direct and standard experimental implementation of these systems using superconducting circuits they can be realized by coupling a transmon qubit to meandering coplanar waveguides \cite{Kannan2020}. These transmon qubits are coupled to distant points along the waveguide, so there can be significant phase differences and  time delays of the propagating radiation along the waveguide. This situation is highly non-Markovian and gives rise to revivals of the emitter population, frequency-dependent rates, and to the creation of bound states inside the continuum (BIC). BIC states arise from interference and have being experimentally observed in a broad variety of physical systems \cite{AzzamAOM2021}.  A very interesting result is the ability of a giant atom to create persistently oscillating bound states as consequence of the overlap of two bound states inside the continuum \cite{KockumPRR2020}, a fact that could find potential applications in the encoding and manipulation of quantum information in cavity-QED.
\\

Inspired by the work of Berman and Ford \cite{BERMAN2010175}, in this paper we study the non-Markovian dynamics of a minimal model for a giant emitter coupled to a cavity-array waveguide beyond the Weisskopf-Wigner approach. Our treatment extends the applicability of the formalism to current interesting and experimentally relevant systems in the context of giant atoms. The paper is organized as follows:
In section \ref{cavity::array::model} we describe the main model employed throughout the paper. It consists of a single emitter coupled to cavity-array waveguide. In section \ref{method::} and \ref{application::cavity::array::model} we describe the general formalism based on the exact inversion of the Laplace transform and its application to the basic model. In section \ref{giant::atom::application} we show that these methods can be extended to the case of giant emitters in waveguide-QED and present an exactly solvable minimal model. In section \ref{circuit::QED::model} a full microscopic derivation of the Hamiltonian based on a lumped-element circuit model is given, and finally in section \ref{conclusions} we outline the conclusions. 
\section{Model}
\label{cavity::array::model}
Let us consider a two-level system with excited state energy $\delta$, coupled to a one-dimensional array of cavities acting as an effective waveguide.
The Hamiltonian of this system reads as follows, 
\begin{equation}
\label{cavity::array::ham}
H = \delta \sigma^{+}\sigma^{-} + \omega_0\sum_{n}a^{\dagger}_n a_n+\xi\sum_{n}\left(a^{\dagger}_{n+1}a_{n}+a^{\dagger}_{n}a_{n+1}\right) +g_0\left(\sigma^{+}a_0+\sigma^{-}a^{\dagger}_{0}\right),
\end{equation}
where $\omega_0$ is the on-site cavity energy and $\xi$ is the hooping parameter between nearest-neighbor cavities. The two-level system couples to the cavity array trough a single cavity located at the center of the waveguide at $x_0=0$.
The cavity-array waveguide can be diagonalized by taking advantage of the translational invariance trough the discrete Fourier transform of the bosonic operators, i.e., $a_k =N^{-1/2}\sum_{n}e^{ikn} a_n$. Applying this transformation to the full Hamiltonian \eqref{cavity::array::ham} and moving to a rotating frame with respect to the natural frequency of the cavities $\omega_0$, we obtain,
\begin{eqnarray}
    H = \Delta\sigma^{+}\sigma^{-}+\sum_{k}\omega_k a^{\dagger}_{k}a_{k}+\frac{g_0}{\sqrt{N}}\sum_{k}\left(\sigma^+ a_{k}+\sigma^{-}a^{\dagger}_{k}\right),
\end{eqnarray}
where $\Delta=\delta-\omega_0$, and $\omega_k=2\xi\cos k$ is the dispersion relation of the cavity-array waveguide. 
An initially excited two-level system will emit radiation into the waveguide modes as a consequence of the spontaneous emission. In the single-excitation manifold we can describe this process by computing the probability amplitudes associated to finding the atom in the excited sate $\ket{e}$ and the cavity array in the collective vacuum $\ket{\mathbf{0}}=\ket{0...0}$, which we call $\alpha$, and the one corresponding to the atom in the ground state $\ket{g}$ and a photon occupying the $k-$th mode $a^{\dagger}_{k}\ket{\bold{0}}$, which we name $\beta_{k}$. Using the Scrh\"odinger equation we get the following coupled differential equations for the probability amplitudes, 
\begin{eqnarray}
\label{coupled::eqs}
i\dot{\tilde{\alpha}}&=&\sum_{k}g_{k}\tilde{\beta}_{k}e^{-i(\omega_k -\Delta)t}, \nonumber \\
    i\dot{\tilde{\beta}}_{k}&=& g_{k}\tilde{\alpha} e^{i(\omega_k -\Delta)t},
\end{eqnarray}
where we have made the following change of variables: $\tilde{\alpha}=e^{i\Delta t}\alpha$ and $\tilde{\beta}_{k}=e^{i\omega_k t}\beta_k$. We can decouple eqs.\eqref{coupled::eqs} and write down the following integro-differential equation for the atomic probability amplitude as
\begin{eqnarray}
\label{integro::diff::eq}
\dot{\tilde{\alpha}}(t)=
    -\frac{1}{2\pi}\int_{0}^{t}d\tau \int_{\omega_{\rm{min}}}^{\omega_{\rm{max}}}d\omega J(\omega)e^{-i(\omega-\Delta)\tau}\tilde{\alpha}(t-\tau),
\end{eqnarray}
where we have introduced the spectral function $J(\omega)\equiv2\pi\sum_{k} |g_k|^2 \delta(\omega-\omega_k)$, that characterizes the effects of the bath of harmonic oscillators in Fourier space on the dynamics of the two-level system \cite{Weiss2012,IdeVega2017}. Here $g_k=g_0/\sqrt{N}$, which is independent of $k$.
We can explicitly compute the spectral function in the continuum limit by taking $\sum_k\to (N/2\pi)\int^{\pi}_{-\pi} dk$ in the definition of the spectral function. This leads to the following expression \cite{IdeVega2017} 
\begin{eqnarray}
\label{spectral::function::cavity::array}
    J(\omega)=2 N |g_k(\omega)|^2 \left|\frac{\partial\omega_k}{\partial k}\right|^{-1},
\end{eqnarray}
provided that the dispersion relation $\omega_k$ has a single branch (see inset figure in fig.\ref{fig::1}(b)). The spectral function for this case reads, 
\begin{eqnarray}   \label{spectral::function::cavity::array}
    J(\omega)= \frac{2 g^2_0}{\sqrt{4\xi^2-\omega^2}}.
\end{eqnarray}

In fig.\ref{fig::1}(b) we show the behavior  of the spectral function as a function of the frequency. This spectral function has singularities at the band-edge frequencies $\pm2\xi$, and it is almost constant for most frequencies near the center of the band dispersion. As we can see from the expression in eq.\eqref{integro::diff::eq}, near the center of the band the function $J(\omega)$ varies slowly with frequency and we have that $J(\omega)\approx J(0)$. We can then write the usual purely exponential decay expression for the atomic probability amplitude as
\begin{eqnarray}
    \alpha(t)=e^{-(\gamma/2+i\Delta)t},
\end{eqnarray}
being $\gamma=J(0)=g_0^2/\xi$ the decay rate in the Markovian limit. This approximation is obviously not valid for frequencies close to the band edges, and we must use an exact treatment in order to describe the actual dynamics. 
\section{Exact dynamics for a general spectral function}
\label{method::}
\begin{figure}
    \centering
    \includegraphics[width=0.8\linewidth]{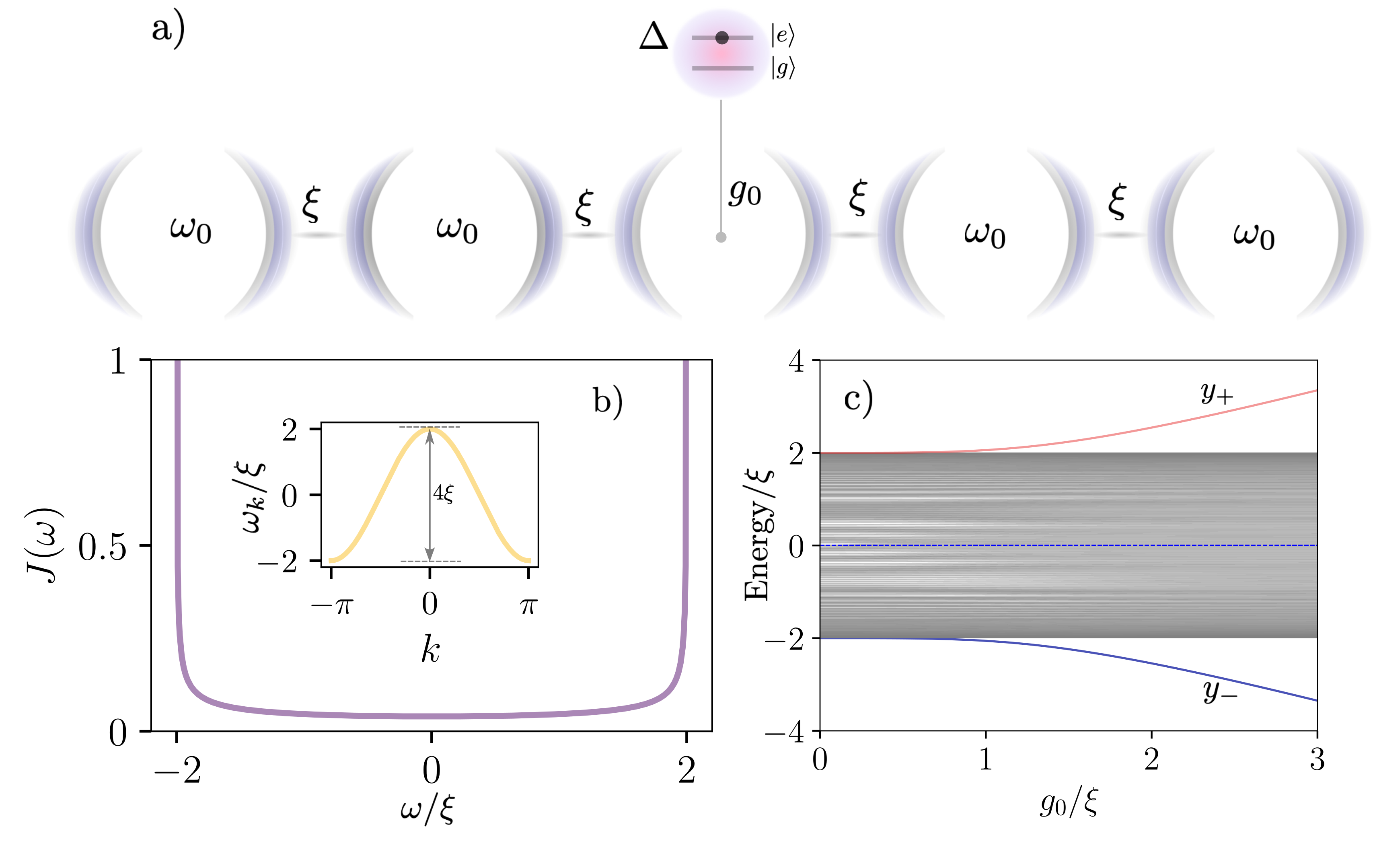}
    \caption{(a) Schematic picture of a two-level system coupled to a cavity-array waveguide at the central cavity. (b) Spectral function given by eq.\eqref{spectral::function::cavity::array} versus frequency. (c) Energy spectrum of the model versus the normalized coupling. The lines above and bellow the continuum are bound states with energies $y_{\pm}$. The coupling value is set to $g_0=\xi/5$ for the rest of figures in the paper.} 
    \label{fig::1}
\end{figure}

In this section we briefly review the method employed by Berman and Ford \cite{BERMAN2010175} to characterize the dynamics of decay of a two-level system into a continuum of modes. This method allows us to find explicit solutions for the probability amplitudes that go beyond the Wigner-Weisskopf approximation, providing a systematic way of dealing with arbitrary spectral functions. This is specially well suited for the study of non-Markovian or memory effects naturally present in the study of structured baths described by nontrivial spectral functions. 
Eq.\eqref{integro::diff::eq} can be formally solved using the Laplace transform. By doing so and taking the formal inverse Laplace transform we can write the formal solution as
\begin{eqnarray}
\label{bronwich::integral}
    \tilde{\alpha}(t)=\frac{e^{i\Delta t}}{2\pi i}\int_{\sigma-i\infty}^{\sigma+i\infty}ds\frac{e^{st}}{s+i\Delta+G(s)},
\end{eqnarray}
where we have defined the function 
\begin{eqnarray}
    G(s) = \frac{1}{2\pi}\int_{\omega_{\rm{min}}} ^{\omega_{\rm{max}}}d\omega\frac{J(\omega)}{s+i\omega}.
\end{eqnarray}
It can be shown that the function $G(s)$ has a discontinuous jump along the imaginary axis, from $-\omega_{\rm{max}}$ to $-\omega_{\rm{min}}$, i.e, there exists a branch cut along this interval.
Choosing the integration contour depicted in fig.\ref{fig2}(a) and applying the residue theorem we arrive at the following exact solution
\begin{eqnarray}
\label{exact::general::solution}
    \alpha(t)&=&\lim_{\epsilon\to\ 0}\int_{-\omega_{\rm{max}}}^{-\omega_{\rm{min}}}\frac{dy}{2\pi} \frac{J(-y) e^{i y t}}{\left[y+\Delta-iG(-\epsilon+i y)\right]\left[y+\Delta-iG(\epsilon+i y)\right]} \\ \nonumber&+&\sum_{j}r_{j} e^{-i y_j t},
\end{eqnarray}
where $r_j=\left[1+G'(s)\big{|}_{s=-iy_j}\right]^{-1}$, with the poles $y_j$ being solutions to the equation
\begin{eqnarray}
\label{poles::eq}
    y-\Delta+iG(-iy)=0. 
\end{eqnarray}
The existence of such poles depends entirely on the function $J(\omega)$. In the case of the cavity array treated here there are two poles that lie outside the branch cut, bellow and above the branch points. The solution in eq.\eqref{exact::general::solution} shows two contributions. The first term represents the contribution from scattering states inside the continuum (fig.\ref{fig::1} (c)). In the long time limit this term decays to zero and  does not contribute to the steady state. The second term however, obtained from the residues, gives rise to a finite contribution in the steady state. This is always the case if the spectral function $J(\omega)$ is bounded, either from above or bellow \cite{BERMAN2010175}. In the case of the the spectral function given in \eqref{spectral::function::cavity::array} there are two bound states, one bellow and one above of the continuum of states, corresponding to two distinct solutions of the equation of eq.\eqref{poles::eq}. The energy of these states is plotted in fig.\ref{fig::1}(c) as a function of the coupling where we can see that they move  farther away from the continuum as we increase the coupling to the waveguide. In the following section we illustrate the use of this general method and solve the dynamics for the cavity array bath and its spectral function exactly. 
%We should mention that this formalism can be applied to any spectral function  
%For the rest of this manuscript we will assume to have $\omega_0=0$, as it facilitates calculations and does not compromise generality. 

\section{Exact solution for a cavity-array waveguide}
\label{application::cavity::array::model}
\begin{figure}
    \centering
    \includegraphics[width=1.0\linewidth]{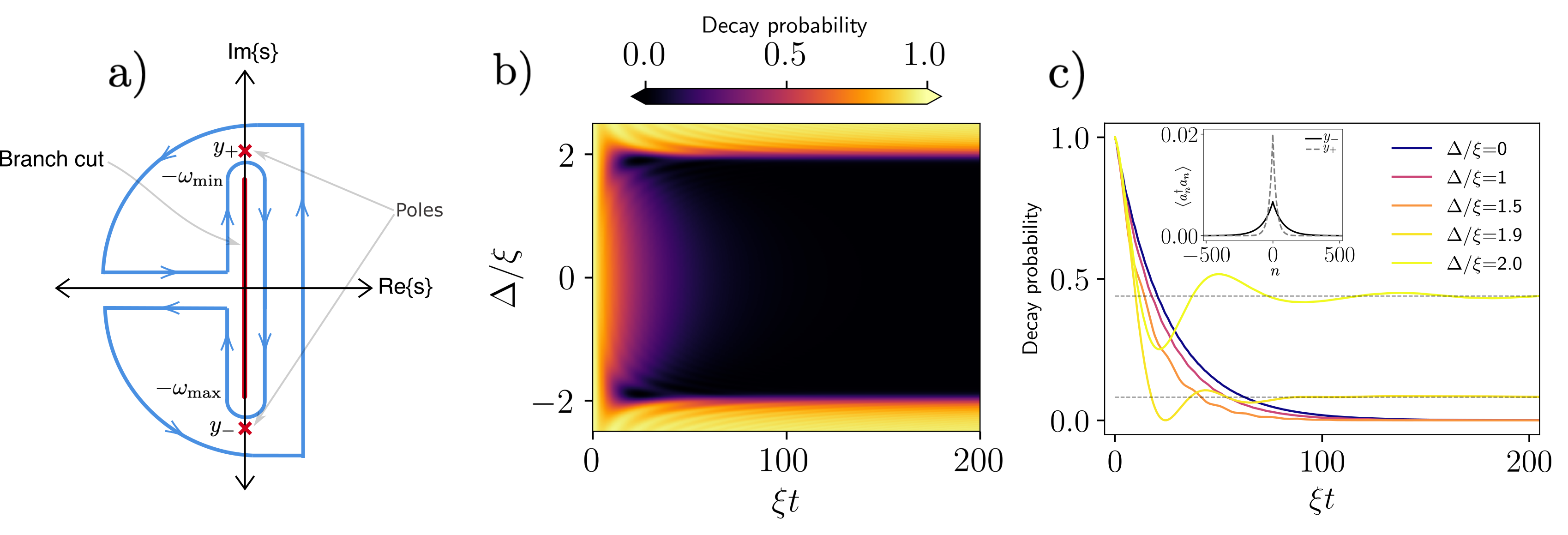}
    \caption{(a) Contour integration path in the complex plane. (b) Evolution of the excited state probability along the frequency band. (c) Representative cases of excited state dynamics. The inset in (c) shows the exponentially localized photonic state for both bound states and $\Delta=3\xi/2$.}
    \label{fig2}
\end{figure}

In order to apply the formal solution in \eqref{exact::general::solution} we employ the explicit form of the spectral function given in eq.\eqref{spectral::function::cavity::array}, together with the expression for $G(s)$ in this case, which reads
\begin{eqnarray}
    G(s)= \frac{g_0^2}{s\sqrt{1+4\xi^2/s^2}}. 
\end{eqnarray}
As mentioned above this function has a branch cut along the imaginary axis, within the interval $[-2\xi,2\xi]$. Using the fact that 
\begin{eqnarray}
    \lim_{\epsilon\to 0}  G(\pm\epsilon+i y)=\pm \frac{g_0^2}{\sqrt{4\xi^2-y^2}},
\end{eqnarray}
we can obtain the following expression for the excited state atomic probability amplitude, 
\begin{eqnarray}
    \alpha(t)= \frac{4 g_0 ^2}{\pi\xi^2}\int_{-1}^{1}d\tilde{y}K(\tilde{y})e^{2i\xi\tilde{y}t} + r_{+} e^{-i y_{+} t}+ r_{-} e^{-i y_{-} t},
\end{eqnarray}
where the integral kernel is given by,
\begin{eqnarray}
    \label{kernel::small}
    K(\tilde{y})=\frac{\sqrt{1-\tilde{y}^2}}{4(2\tilde{y}+\Delta/\xi)^2(1-\tilde{y}^2)+(g_0/\xi)^4},
\end{eqnarray}
 and the residues
 %{\color{red}
 %NEED TO FIX THE $\omega_0$ issue here}
 \begin{eqnarray}
     r_{\pm}= \left[1+\frac{g_0^2}{y^2_{\pm}(1-4\xi^2/y^2_{\pm})^{3/2}}\right]^{-1}.
 \end{eqnarray}
 Notice that we have made the change of variable $\tilde{y}=y/2\xi$. The poles $y_{\pm}$ are real solutions to eq.\eqref{poles::eq} and represent the energy of two bound states outside the continuum band. This solution has been equivalently found in a different way in Ref.\cite{SanchezBurilloPRA2017} by spanning the state vector into bound and scattering eigenstates. One nice thing about the present method is that there is no need to separate the contributions coming from the continuum and the bound modes, since they appear naturally as consequence of the integration around the branch cut and the residues of each pole inside the contour, respectively.  In fig.\ref{fig2}(b) we show the decay probability of the two-level system as a function of time varying the qubit frequency from $-5\xi/2$ to $5\xi/2$. We can see that if the emitter is on resonance (or close to) with the center of the band the dynamics is indeed very well described by an exponential decay, corroborating the Markovian approximation. Oscillations in the decay probability start to appear as we get detuned from resonance. Close to the band-edges we start to gain some contribution from the bound state (the state corresponding to $y_+$ in this case), and the decay becomes partial. This means that part of the excitation is trapped around the position of the emitter, a phenomenon that is well known in this type of systems and has been well studied in the literature of waveguide-QED \cite{CalajoPRA2016,SanchezBurilloPRA2017}.  The non-zero asymptotic value of the decay probability is given exactly by the square of the residue $r^2_+$ (dotted lines in the main plot of fig.\ref{fig2}(c)). Fig.\ref{fig2}(c) also shows the exponentially localized wave functions in position space of the radiation field and their asymmetry as we move towards the upper band limit. In this case the value of the positive pole (energy of the upper bound state) increases as well as the localization length of its wave function.
%{\color{red}  mostrar algunos plots interesantes, no se que tanto decir acerca de bound states que no este ya dicho en la literatura... No se si mencinar mas cosas del Kernel y sus propiedades, quiza mejor en el atomo gigante

%Mencionar la asimetria de BS cuando movemos $\Delta$
%\\
%to do: hacer las graficas del campo (dos) y tambien mencionar lo relevante. creo que con esto acabo la primera parte del manuscrito.
%DISCUTIR LAS FIGURAS Y CREO QUE YA ESTARIA}.
\begin{figure}
    \centering
    \includegraphics[width=0.7\linewidth]{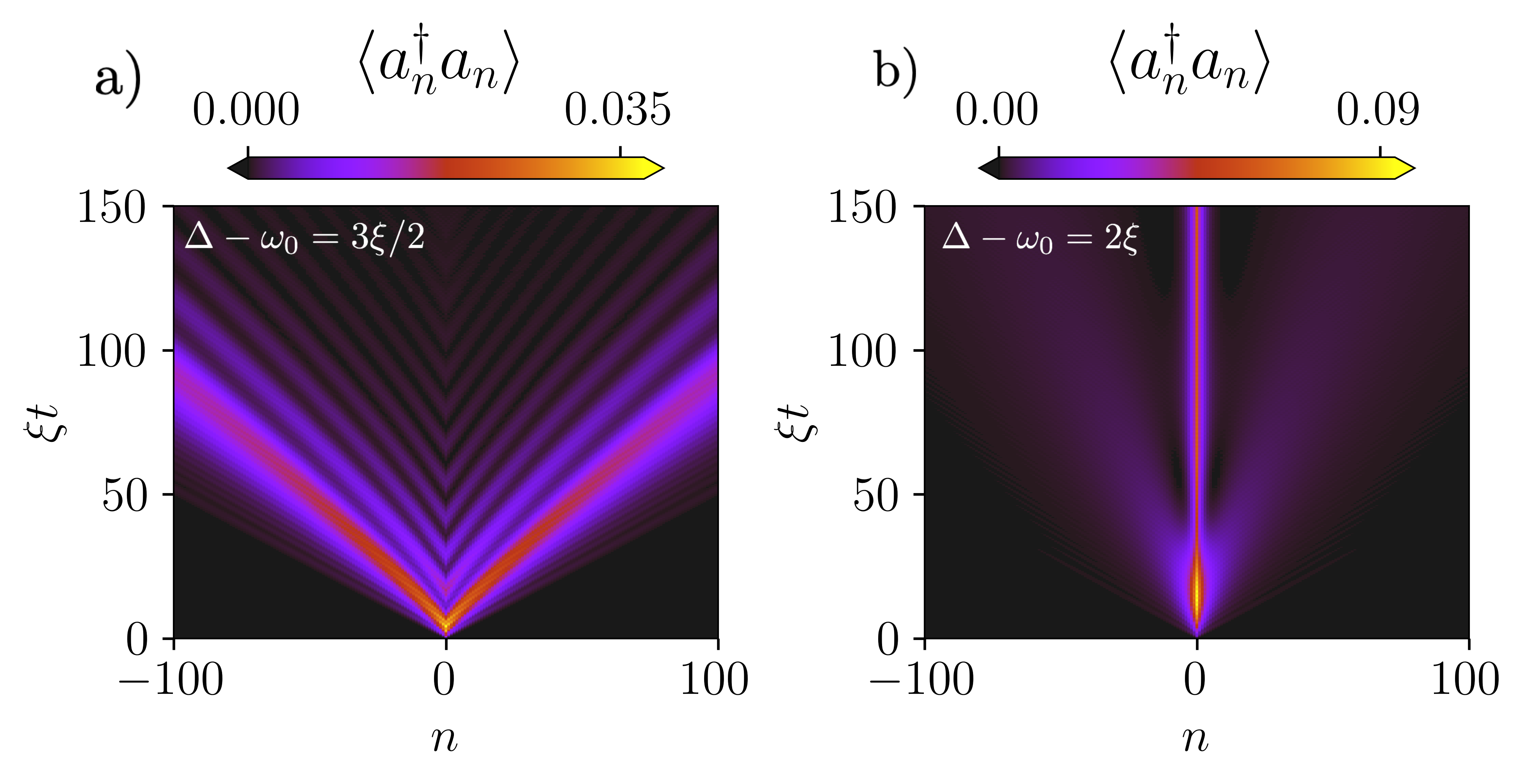}
    \caption{Time evolution of the radiation field along the cavity-array waveguide for a propagating case with (a) $\Delta=3\xi/2$ and for a bound state with (b) $\Delta=2\xi$.}
    \label{fig::3}
\end{figure}
The probability amplitude of the radiation field in position space can be obtained from the Fourier transform of the one in momentum space, i.e., $\beta_n=(1/\sqrt{N})\sum_k\beta_{k}e^{-ik n}$. Using the expression for $\beta_k$ in eq.(\ref{coupled::eqs}) we can write,
\begin{eqnarray}
    \beta_{n}(t)=-i^{|n|+1}g_0 \int_{0}^{t} d\tau\alpha(t-\tau)e^{-i\omega_0\tau}J_{|n|}(2\xi\tau).
\end{eqnarray}
For values of $\Delta$ closer to the upper/lower band limit it is possible to compute the steady state of the radiation field as,
\begin{eqnarray}
     \lim_{t\to\infty}|\beta^{\pm}_n(t)|^2=\expval{a^{\dagger}_{n}a_{n}}_{\pm}= \frac{g^2_0 r^2_{\pm}}{y^2_{\pm}-4\xi^2}e^{-2\kappa_{\pm}\abs{n}},
\end{eqnarray}
where $\kappa_{\pm}=\lambda_{\pm}^{-1}=\cosh^{-1}\left({|y_\pm|/2\xi}\right)$, being $\lambda_{\pm}$ the corresponding localization length for each bound state, respectively. For values of $\Delta$ not that close to the upper/lower limit band and off-resonance from the center of the band, there can be a small contribution from the other bound state. In this case there is a small interference term $\sim r_{+}r_{-}e^{-(\kappa_{+}+\kappa_-)\abs{n}}\cos[(y_+-y_-)t]$. In fig.\ref{fig::3} we show the dynamics of propagation of the radiation field for two relevant cases. The first situation is when the frequency of the emitter lies inside the band at $\Delta=3\xi/2$. This corresponds to a field that propagates out of the origin at a speed given by the group velocity evaluated at the corresponding value of the momentum $k$. The second case illustrates the dynamical formation of the upper bound state that localizes exponentially around the emitter as we can observe from fig.\ref{fig::3}(b).

\section{A single giant emitter in a cavity-array waveguide}
\label{giant::atom::application}
\begin{figure}
    \centering
    \includegraphics[width=0.8\linewidth]{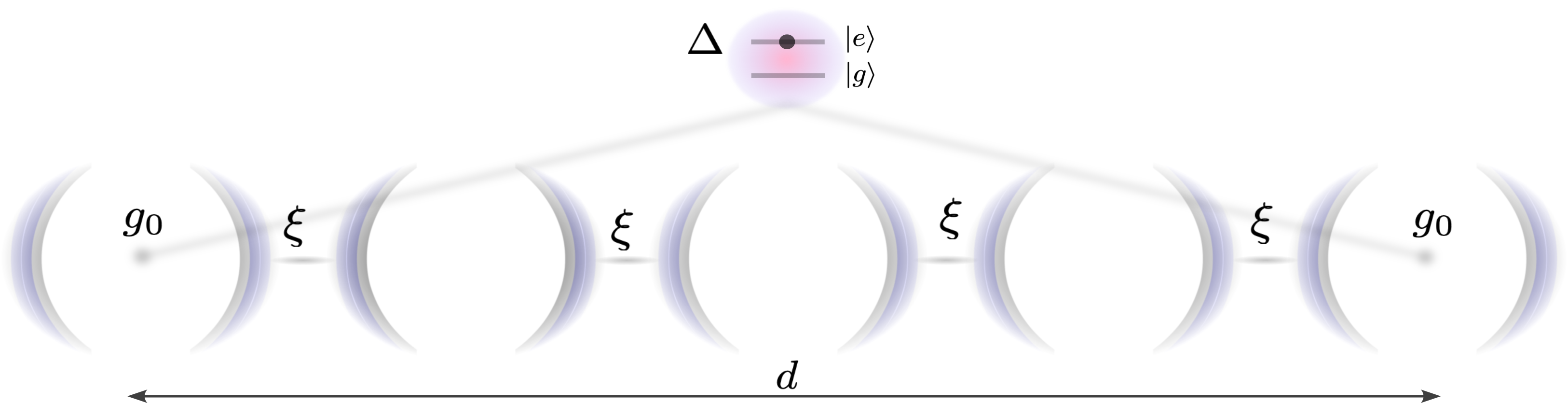}
    \caption{Minimal model for a giant emitter coupled to two distant cavities ($N_c=2$).}
    \label{fig::4}
\end{figure}

Let us now consider a minimal model for a giant emitter. This consists of a two-level system coupled to $N_c$ cavities at positions $x_j$, as we illustrate in fig.\ref{fig::4} for $N_c=2$. The Hamiltonian for this model is a straightforward generalization of the Hamiltonian \eqref{cavity::array::ham}, and it is given by,
\begin{eqnarray}
\label{Hamiltonian::giant::atom}
H = \delta \sigma^{+}\sigma^{-} &+& \omega_0\sum_{n}a^{\dagger}_n a_n+\xi\sum_{n}\left(a^{\dagger}_{n+1}a_{n}+a^{\dagger}_{n}a_{n+1}\right) \\
&+&g_0\sum^{N_c}_{j=1}\left(\sigma^{+}a_{x_j}+\sigma^{-}a^{\dagger}_{x_j}\right). \nonumber
\end{eqnarray}
Transforming to the Fourier space and assuming equally symmetric spaced couplings with respect to the position of the emitter we get,
\begin{eqnarray}
    H = \Delta\sigma^{+}\sigma^{-}+\sum_{k}\omega_k a^{\dagger}_{k}a_{k}+\sum_{k}\tilde{g}_{k}\left(\sigma^+ a_{k}+\sigma^{-}a^{\dagger}_{k}\right),
\end{eqnarray}
where the couplings are now momentum-dependent, i.e., 
\begin{eqnarray}
\tilde{g}_{k}=\frac{g_0}{N_c\sqrt{N}}\frac{\sin(kd N_c/2)}{\sin(kd/2)}.
\end{eqnarray}
On can compute the effective spectral function associated to the problem, which reads,
\begin{eqnarray}
    J_{\rm{eff}}(\omega)= J(\omega)\mathcal{G}(\omega), \,\,\,\,\,\,\,\, \,\,\mathcal{G}(\omega)= \frac{1}{N^2_c}\frac{1-\cos(k(\omega)d N_c)}{1-\cos(k(\omega)d)}.
\end{eqnarray}
where $J(\omega)$ is the spectral function corresponding to the emitter coupled to a single cavity in the waveguide (eq.\eqref{spectral::function::cavity::array}), and $\mathcal{G}(\omega)$ is a function of the frequency and the distance between coupling cavities modifying the spectral function as a consequence of considering a giant emitter \cite{TerradasPRA2022}.  In the following we focus on the particular case of $N_c=2$ as a minimal model. This case shows non-trivial dynamics and gives rise to the appearance of bound states inside the continuum (BIC). 

%{\color{red}  It is actually a good idea to start from the most general model and then show solutions for particular cases. So you better write the most general Hamiltonian and spectral function.}
\subsection{Exact solution for $N_c=2$}
In order to explore concrete examples and to obtain explicit expressions for the dynamics of the model with a giant emitter we will focus on the case of $N_c=2$. This is perhaps the simplest model of a giant emitter having only two equidistantly spaced coupling points to the waveguide. 
The effective spectral function can be written as,
\begin{eqnarray}
    J_{\rm{eff}}(\omega)=\frac{g_0^2}{\sqrt{4\xi^2-\omega^2}}\left[1+\cos(k(\omega)d)\right].
\end{eqnarray}
\begin{figure}
    \centering
    \includegraphics[width=0.8\linewidth]{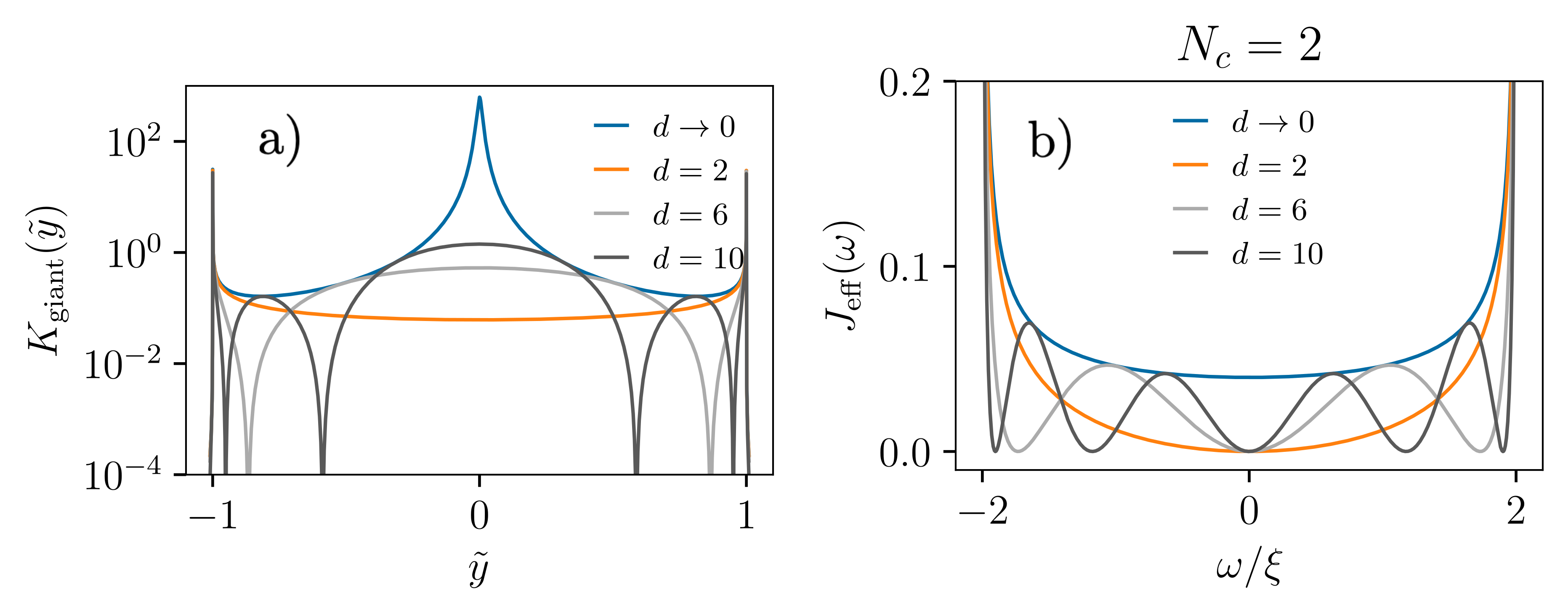}
    \caption{(a) Integral kernel in eq.\eqref{kernel::giant} along the frequency band. (b) Effective spectral function for different separation distances.}
\label{fig5::spectral::function::kernel}
\end{figure}
Interestingly, this spectral function has $\lceil d/2 \rceil$ zeros at specific frequencies given by $\omega_m=2\xi\cos(\pi(2m+1)/d),$ with $m\in\mathbb{Z}$. This is shown in fig.\ref{fig5::spectral::function::kernel}(b) where we show the effective spectral function as a function of frequency for different separation distances $d$. Notice that the function $G(s,d)$ is distance-dependent in this case, and can be obtained in a closed form as, 
\begin{eqnarray}
    G(s,d) = \frac{g^2_0}{2}\frac{{\sign(\Re\{s}\})}{s\sqrt{1+4\xi^2/s^2}}\left[1+\varrho^{d}\right], 
\end{eqnarray}
where $\varrho(s)= is/2\xi-i\sign(\Re\{s\})\sqrt{1+s^2/4\xi^2}$. The discontinuous jump across the branch cut is given by the difference, 
\begin{eqnarray}
    \lim_{\epsilon\to 0}G(\epsilon+iy, d)-\lim_{\epsilon\to 0}G(-\epsilon+iy, d)&=& \frac{g_0^2}{\sqrt{4\xi^2-y^2}}\left[1+\cos(k(-y)d)\right]\nonumber\\
    &=&J_{\rm{eff}}(-y).
\end{eqnarray}
We can then write the solution as
\begin{eqnarray}
    \alpha(t) = \frac{4g_0^2}{\pi\xi^2}\int^1_{-1}d\tilde{y}K_{\rm{giant}}(\tilde{y},d)e^{i2\xi\tilde{y}t}+\sum_{j={\pm}}r_j e^{-i y_j t}, 
\end{eqnarray}
being the kernel function
\begin{eqnarray}
\label{kernel::giant}
    K_{\rm{giant}}(\tilde{y},d)= \frac{\sqrt{1-\tilde{y}^2}\left[1+\cos(k(-\tilde{y})d)\right]}{8(2\tilde{y}+\Delta/\xi)^2(1-\tilde{y}^2)+F(\tilde{y},d)}, 
\end{eqnarray}
with $F(\tilde{y}, d)= (g_0/\xi)^4 (1+\cos(k(-\tilde{y})d))-4(g_0/\xi)^2(2\tilde{y}+\Delta/\xi)\sqrt{1-\tilde{y}^2}\sin(k(-y)d).$ 
Let us first discuss the contribution coming from scattering states to the dynamics. 
In the limit of $d\to 0$ the above kernel simplifies to the kernel for a small emitter given in eq.\eqref{kernel::small}. 
%{\color{red} CHECK OUT THE RIEMANN-LEBESGUE LEMMA. IT SEEMS TO NOT APPLY FOR THE GIANT ATOM KERNEL!}
%{\color{red} Plot the kernel and verify this is the correct expression for it. you can compare the behavior of both Kernel functions and say something about it}
It is interesting to analyze the behavior of the kernel function as it encodes information about the dynamics. For instance, the sharpest peak in this kernel will dominate the long-time dynamics. This is the case for the small emitter, where the narrow peaks close to the band-edges give rise to a power law decay $\sim t^{-1}$ of the occupation probability at intermediate times, that eventually transitions to a faster algebraic decay $\sim t^{-3}$ at very long times  \cite{SanchezBurilloPRA2017}. Let us first describe the situation of $\Delta$ near to the center of the band. In fig.\ref{fig5::spectral::function::kernel}(a) we plot the kernel function given in eq.\eqref{kernel::giant} along the frequency band. For $d\to 0$ (small emitter) we see that there is a main peak centered at the middle of the band and two singularities at the band-edges located at $\tilde{y}=\pm1$. Near to these singularities we find two local maxima very close to $\pm1$. As the emitter gets giant the kernel shows an interesting behavior. For example, for $d=2(2n+1)$ there is no central peak ($F(0,d)=0$) and the kernel behaves smoothly around the center of the band. 
The kernel function has $\lceil d/2 \rceil-1$ zeros
along the band located at same frequencies that the ones for the effective spectral function, except for the one at $\tilde{y}=0$, since the kernel function has a finite value at the center of the band. For large distances $d$,  more zeros in the kernel function start to appear, making the kernel a highly oscillatory function. Contrary to the case of a small emitter, 
the initial decay is no longer well approximated by an exponential behavior and the contribution from scattering states gets enhanced as the distance is increased. This behavior is expected due to the fact that there is no Lorentzian-like central peak.
For distances such that $d=4n$ the central peak for the small emitter case survives but becomes rounded with faster decaying tails for large $n$. In this case we obtain an approximately initial exponential decay with a rate of $\sim\gamma/2$. Interestingly, after a time $ t=d/2\xi$ this decay gets enhanced or accelerated as consequence of the constructive interference of the two initial emitted radiation wavepackets from each distant cavity. The behavior of the decay probability
at long times remains similar to the small emitter situation showing a power law behavior $\sim t^{-3}$.

\subsection{Bound states inside the continuum}
\begin{figure}
    \centering
    \includegraphics[width=1.0\linewidth]{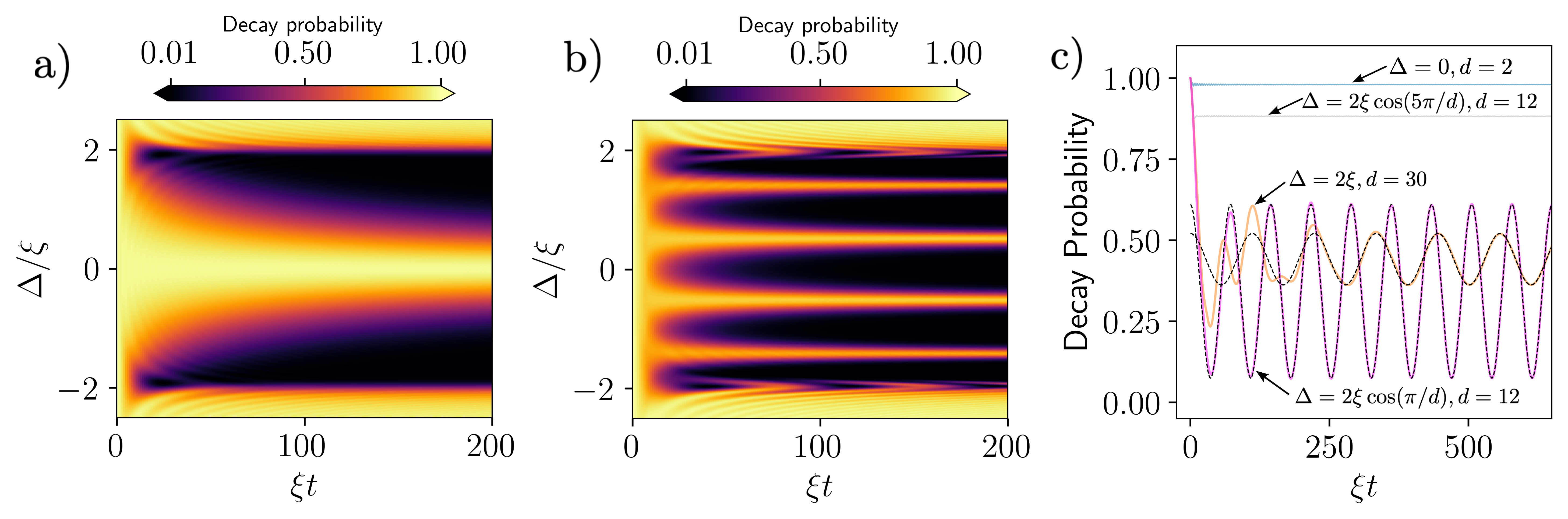}
    \caption{Evolution of the excited state probability along the frequency band for (a) $d=2$ and (b) $d=12$. (c) Evolution of BIC states for $\{\Delta=0, d=2\}$ and $\{\Delta=2\xi\cos(5\pi/12), d=12\}$.}
    \label{fig:6}
\end{figure}

\begin{figure}
    \centering
\includegraphics[width=1.025\linewidth]{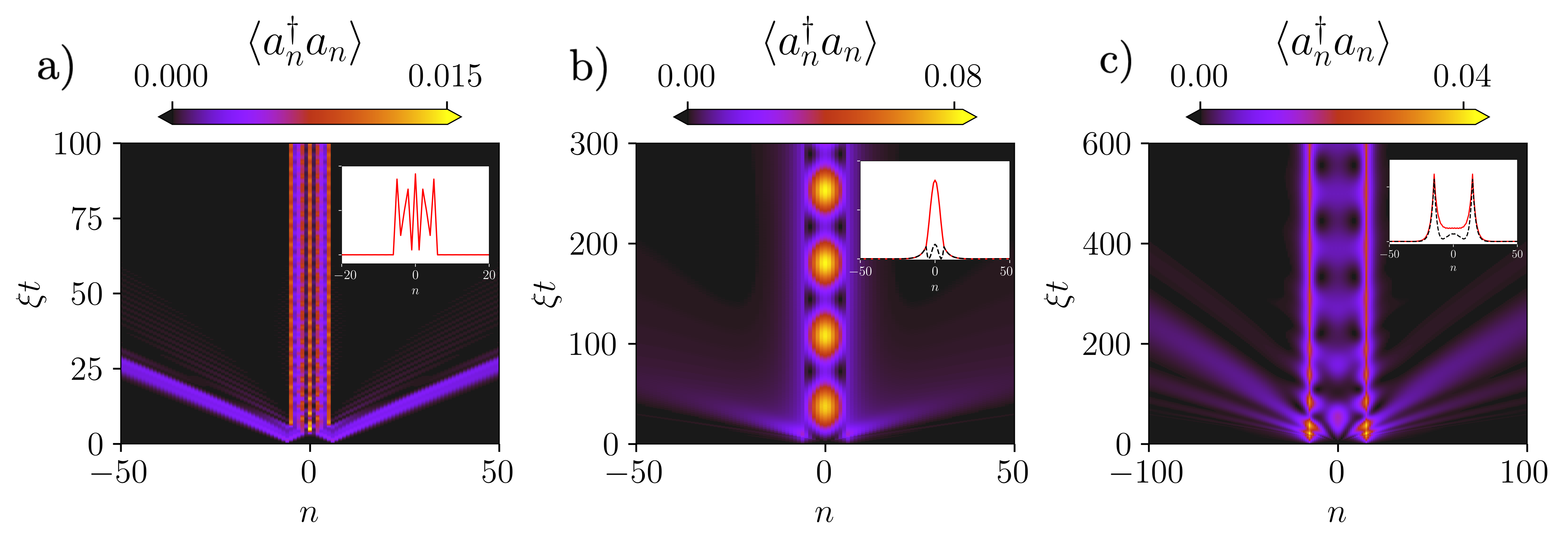}
    \caption{Time evolution of the radiation field along the cavity-array waveguide for (a) BIC state with same parameters as in fig.\ref{fig:6}(c) for $\{\Delta=2\xi\cos(5\pi/12), d=12\}$. (b) Oscillating bound state between BOC and BIC $\{\Delta=2\xi\cos(\pi/d), d=12\}$. (c) quasi-oscillating bound state for $\{\Delta=2\xi, d=30\}$. Inset plots in each figure show the stationary field profile of the BIC state in case (a), and the maximum and minimum intensity profiles for the (b) oscillating and (c) quasi-oscillating fields.} 
    \label{fig:7}. 
\end{figure}

So far we have discussed the dynamics in terms of scattering and bound states that can exist outside the continuum of the allowed modes. It turns out that bound states can also be found inside the continuum and we refer to them as BIC states \cite{LonghiEPJB2007}. 
In order to search for the existence of such states in our minimal model we need to look at the zeros of the effective spectral function. From our earlier discussion we already encountered that $J_{\rm{eff}}(\omega_m)=0$ for $\omega_m=2\xi\cos(\pi(2m+1)/d).$ If the frequency of emitter is such that $\Delta=\omega_m$ we have that  the equation for the poles is $y-\omega_m+iG(-iy)=0$, which can be satisfied for $y=\omega_m$ since $G(-i\omega_m)=0.$ Therefore, if the frequency of the emitter coincides with a zero of the effective spectral function there will be a non-zero contribution from a pole of the emitter probability amplitude $\alpha(s)$ in the Laplace domain \cite{LonghiEPJB2007}. We can then describe the dynamics in this situation (provided that there is no contribution from the bound states outside the continuum) as following,
\begin{eqnarray}
\label{alpha::BIC}
     \alpha(t) = \frac{4g_0^2}{\pi\xi^2}\int^1_{-1}d\tilde{y}K_{\rm{giant}}(\tilde{y},d)e^{i2\xi\tilde{y}t}+r_m e^{-i \omega_m t},
\end{eqnarray}
where $r_m=\left[1+G'(s,d)\big{|}_{s=-i\omega_m}\right]^{-1}$, is the residue corresponding to the pole at $s=-i\omega_m$. An exact expression for this residue for arbitrary $d$ is shown in the appendix \ref{appendixA}.
%{\color{red} CALCULAR EL CAMPO ANALITICAMENTE PARA EL GIANT ATOM. Creo que es mejor poner dos figuras, una para $d=2$ y otra para $d$ grande. La ultima figura del panel debe ilustrar casos genericos y asi podemos discutirlos en el texto.
%Need to find out if the state I think is not decaying it is truly not decaying}
The situation of a giant emitter coupled simultaneously to two cavities is probably a minimal model for the existence of a bound state inside the continuum. A BIC state arises for $d=2$ when the frequency of the emitter is tuned exactly at the center of the band, as shown in figures \ref{fig:6}(a) and  \ref{fig:6}(c). We see that there is fast initial decay given by the contribution from the scattering states followed by the contribution from the residue of the corresponding BIC, being this the dominant term resulting in a nonzero steady state probability at long times. As we increase $d$ more BIC states are allow to exists inside the continuum as we can notice from fig.\ref{fig:6}(b) for $d=12$. The corresponding field profile for a BIC state can be obtained exactly in the long time limit and it is given by,
\begin{eqnarray}
\expval{a_{n}^{\dagger}a_n}_{\text{\tiny{BIC}}} =\frac{g_0^2 r^2_m}{2(4\xi^2-\omega^2_m)}\left[1+\cos(\phi(|n+d/2|-|n-d/2|))\right],  
\end{eqnarray}
where $\phi=\arcsin(\Delta/2\xi)+\pi/2$.
%{\color{red} 
%plot three figures with the field, for $d=12$, (bic and bic+boc) and quasibic.} 
Figure \ref{fig:7}(a) shows the formation of a BIC state and its profile at long times. Notice that the field is confined over the effective length of the giant emitter, i.e. between the two distant cavities coupling to the emitter, and is zero outside this region due to the destructive interference. This confinement of radiation in which the giant emitter behaves an an effective cavity has also been found in the case of two distant emitters as a consequence of the formation of BIC states \cite{SinhaPRL2020}.  
\subsection{Coherent oscillations of bound states}
Interestingly, the giant emitter model also allows the coherent superposition of two bound states of different type, inside and outside the resonant band. This persistent coherent oscillation has also been found for BIC or dark states in the continuum of Ohmic waveguides for a giant atom having at least $N_c=3$ simultaneous equidistant interactions  \cite{KockumPRR2020}. Here, however, we show that it is possible to have a coherent superposition of two distinct bound states lying outside and inside the continuum. Such sates take place whenever the frequency of the emitter satisfies the condition $J_{\rm{eff}}(\omega_m)=0$, while simultaneously lying sufficiently close to the band edge in order to allow the contribution from a pole outside the band. Under such conditions we have that the long time limit of the occupation probability is given by 
\begin{eqnarray}
    \label{bic+boc::formula}
 \lim_{t\to\infty}|\alpha(t)|^2=r^2_{+}+r^{2}_{m}+2r_{+}r_{m}\cos(y_{+}-\omega_m)t,
\end{eqnarray}
where we have chosen the emitter frequency to match the frequency of the BIC state at $\Delta=\omega_m$ very close the upper band limit. This dynamics at long times is illustrated in fig.\ref{fig:6}(c) for $d=12$, where a non-decaying oscillating bound state with large amplitude is shown. We can see that eq.\eqref{bic+boc::formula} agrees perfectly with the oscillation at long times for $\Delta=\omega_0=2\xi\cos(\pi/12).$ 
Another interesting situation is shown to happen for frequencies of the emitter non exactly matching the BIC condition and also lying sufficiently close of the band edge. In this case it is also possible to generate coherent Rabi-like oscillations lasting for very long times. These states will eventually decay since they are not truly bound states but quasi-bound states. We show one example of such state in fig.\ref{fig:6}(c) and its corresponding field in fig.\ref{fig:7}(c) for $d=30$ and $\Delta=2\xi$. In this case the integral over scattering states decays very slowly, having a quasi-steady state value for a long period of time before decaying algebraically at very long times. Strictly speaking, these states are coherent superpositions of slow decaying scattering modes and bound states outside the continuum. These oscillations will die at long times but it can be useful to address and have access to such modes in some particular experimental situations.
%{\color{red}
%compute the oscillating field for bic + boc
%}
The field corresponding to the oscillating bound state formed by the superposition of both, the BIC and BOC state is given by,
\begin{eqnarray}
    \expval{a_{n}^{\dagger}a_n}_{\text{\tiny{BIC,BOC}}} = \frac{g^2_0}{4}\left|\frac{r_{+}e^{-i y_{+}t}}{\sqrt{y^2_{+}-4\xi^2}}\Phi_{+}(n)+i\frac{r_m e^{-i\omega_m t}}{\sqrt{4\xi^2-\omega^2_m}}\Phi(n)\right|^2,
\end{eqnarray}
where the field profiles for the two bound states are given by $\Phi_{+}(n)=e^{-(\kappa_{+}-i\pi/2)|n-d/2|}+e^{-(\kappa_{+}-i\pi/2)|n+d/2|}$ and $\Phi(n)=e^{i\phi|n-d/2|}+e^{i\phi|n+d/2|}$.
This is shown in fig.\ref{fig:7}(b) for the evolution of the radiation field creating a breathing radiation mode characterizing an oscillating bound state as a combination of two bound states, one inside and one outside the continuum. The quasi-oscillating bound state of the field is shown in fig.\ref{fig:7}(c).
The possibility to create or engineer such infinite oscillating states can be potentially exploited for the design and tailoring of non-stationary cavities supporting particular modes, either for trapping or manipulation, by means of the natural drive field of an oscillating mode,  the quantum state of additional qubits interacting with the waveguide.    
%\subsection{$N_c>2$}
%{\color{red}
%Can we generalize the kernel for arbitrary $N_c$??
%You need to talk about BICs as zeros of the general case of the spectral function, and show some representative cases for dynamics. NEED TO CHECK INTERFERENCE OF BICS} 
%Our previous analysis can be extended to allow arbitrary number of coupled cavities to the emitter. Of course the expressions for the integration kernel become more involve and   
\subsection{Master equation and entanglement entropy}
We can generalize above results in order to allow the giant emitter to be initialized in any arbitrary initial state having coherences. The giant emitter density matrix at any time can be written as,
\begin{eqnarray}
\label{qubit::density::matrix}
    \varrho(t) =
\begin{pmatrix}
\varrho_{ee}(0)|\alpha(t)|^2 & \varrho_{eg}(0)\alpha(t) \\
\varrho_{ge}(0)\alpha^{*}(t) & \varrho_{gg}(0)+ \varrho_{ee}(0)(1-|\alpha(t)|^2),
\end{pmatrix}
\end{eqnarray}
from which we can obtain the following master equation,
\begin{eqnarray}
    \dot\varrho(t)= -i\tilde{\Delta}(t)[\sigma^{+}\sigma^{-},\varrho]+\Gamma(t)\left[\sigma^{-}\varrho\sigma^{+}-\frac{1}{2}\sigma^{+}\sigma^{-}\varrho-\frac{1}{2}\varrho\sigma^{+}\sigma^{-}\right], 
\end{eqnarray}
where $\tilde{\Delta}(t)=-2\Im{\dot\alpha(t)/\alpha(t)}$, and $\Gamma(t)=-2\Re{\dot\alpha(t)/\alpha(t)}$. The above master equation is time-local but in general non-Markovian, since $\Gamma(t)$ can take negative values \cite{AnderssonPRA2014}. 
%{\color{red} Comment about non-Markovianity using this master equation.}
An interesting quantity to analyze is the entanglement entropy resulting from computing the von Neumann entropy of the reduced density matrix \eqref{qubit::density::matrix}, i.e., $S(\varrho)=-\Tr{\varrho\log\varrho}$. In the spontaneous emission process, the standard theory of Weisskopf and Wigner predicts that an initially excited atom undergoes an irreversible decay to the ground state. This means that an initial pure product state is mapped onto another pure final state. The entanglement entropy at the steady state is then zero and no atom-field entanglement remains at long times. As we have discussed in the previous sections, the presence of bound states modifies the steady state of the emitter-waveguide system and therefore some amount of entanglement generated in the process survives in the long time limit. In particular, if the steady state is an oscillating bound state like the one in eq.\eqref{bic+boc::formula}, the entanglement entropy is a time-dependent oscillating function with a frequency given by $y_{+}-\omega_m$.
%\section{Spectrum of emitted field}
\section{Superconducting circuits model}
\label{circuit::QED::model}
Circuit-QED is the current suitable platform for the implementation and  simulation of giant atoms since the first demonstration using superconducting circuits \cite{Kannan2020}. Here we provide the microscopic circuit model derivation of the full Hamiltonian model in  \eqref{Hamiltonian::giant::atom}.
\subsection{Transmission line model for a cavity-array waveguide}
In order to simulate a cavity-array waveguide we consider a chain of capacitively coupled LC resonators. We describe such a system trough the classical Lagrangian, 
\begin{eqnarray}
    \mathcal{L}=\sum^{N}_{n=1}\left[C\left(\dot{\Phi}_{n}^2-\dot{\Phi}_{n}\dot{\Phi}_{n+1}\right)+\frac{C_{0}}{2}\dot{\Phi}_{n}^2-\frac{\Phi_{n}^2}{2L_0}\right].
\end{eqnarray}
We assume that the chain is grounded at the boundary nodes with $\Phi_0=\Phi_{N+1}=0.$ The conjugated charges associated to each node are obtained via,
\begin{eqnarray}
    Q_{n}=\frac{\partial\mathcal{L}}{\partial\dot\Phi_{n}}=C_{\Sigma}\dot\Phi_{n}-C(\dot\Phi_{n-1}+\dot\Phi_{n+1}).
\end{eqnarray}
The classical Lagrangian can also be written in matrix form defining the flux and charge vectors: $\bold{\Phi}=(\Phi_1,\Phi_2,...,\Phi_N)^{T}$, $\bold{Q}=(Q_1,Q_2,...,Q_{N})^{T},$ as follows, 
\begin{eqnarray}
\mathcal{L}=\frac{1}{2}\dot{\bold{\Phi}}^{T}\mathbb{C}\dot{\bold{\Phi}}-\frac{1}{2}\bold{\Phi}^{T}\mathbb{L}^{-1}\bold{\Phi},
\end{eqnarray}
being $\mathbb{C}$ the capacitance matrix with elements $C_{nm}=C_{\Sigma}\delta_{n,n}-C(\delta_{n,m+1}+\delta_{n+1,m}),$ and $\mathbb{L}$ a diagonal  inductance matrix, respectively. Using this form the charge vector can be written as $\bold{Q}=\mathbb{C}\bold{\dot\Phi}$ and the Hamiltonian is obtained by performing the Legendre transform,
\begin{eqnarray}
    \mathcal{H}&=&\sum_{n}Q_{n}\dot\Phi_{n}-\mathcal{{L}}= \frac{1}{2}\bold{Q}^{T}\mathbb{C}^{-1}\bold{Q}+\frac{1}{2L_{0}}\bold{\Phi}^{T}\bold{\Phi}.
\end{eqnarray}
The inverse capacitance matrix can be  well approximated to first order assuming that $C\ll C_0$, resulting in the following approximate Hamiltonian, 
\begin{eqnarray}
    \mathcal{H}\approx\sum_{n}\left[\frac{Q^2_n}{2C_{\Sigma}}+\frac{\Phi^2_{n}}{2L_{0}}+\frac{C}{C^2_{\Sigma}}Q_{n}Q_{n+1}\right].
\end{eqnarray}
Classical canonical variables are now promoted to quantum hermitian flux and charge operators, satisfying the canonical commutation relation $[\Phi_n,Q_{m}]=i\delta_{n,m}$.  In terms of bosonic creation and annihilation operators they read, 
\begin{eqnarray}
    \Phi_n =\sqrt{\frac{1}{2C_{\Sigma}\omega_0}}\left(a_{n}+a^{\dagger}_{n}\right),\,\,\,\, Q_{n}=-i\sqrt{\frac{C_{\Sigma}\omega_0}{2}}\left(a_{n}-a^{\dagger}_{n}\right), 
\end{eqnarray}
for which $[a_{n},a^{\dagger}_{m}]=\delta_{nm}.$ The second quantized Hamiltonian in the rotating wave approximation is then given by
\begin{eqnarray}
    H_{\rm{photons}}= \omega_0\sum_{n}a_{n}a^{\dagger}_{n}+\xi\left(a^{\dagger}_{n+1}a_{n}+a^{\dagger}_{n}a_{n+1}\right),
\end{eqnarray}
where $\omega_0=(L_0C_{\Sigma})^{-1/2}$, and $\xi=\omega_0 C/2C_{\Sigma}$.
\subsection{Coupling to a giant transmon qubit}
A giant transmon circuit is now coupled to the above transmission line at nodes positioned at $x_j$. This can be done by renormalizing the charge of the transmon Hamiltonian. The full Hamiltonian reads,
\begin{eqnarray}
    H = \frac{1}{C_{\Sigma q}}\left(q+\frac{C_g}{C_{\Sigma'}}\sum_{j}Q_{x_j}\right)^2-E_{J}\cos\left(\frac{2\pi\Phi_q}{\Phi_0}\right)+H_{\rm{photons}},
\end{eqnarray}
where $C_g$ is the coupling capacitance and $C_{\Sigma'}=C_{\Sigma}+C_{g}$ is the renormalized capacitance of the transmission line due to the coupling to the transmon. The second-quantized Hamiltonian of the giant transmon coupled to the transmission line up to first order in $C_g/C_{\Sigma'}$ is given by
\begin{equation}
    H = \Delta b^{\dagger}b -\frac{E_C}{2}{b^{\dagger}}^2b^{2}+\omega_0\sum_{n}a_{n}a^{\dagger}_{n}+\xi\left(a^{\dagger}_{n+1}a_{n}+a^{\dagger}_{n}a_{n+1}\right)+g_0\sum_{j}\left(b^{\dagger}a_{x_j}+ba^{\dagger}_{x_j}\right),
\end{equation}
where $\Delta\approx\sqrt{8E_CE_J}$ and $g_0=C_g\sqrt{\Delta\omega_0/C_{\Sigma_q}C_{\Sigma'}}/2.$ Limiting our transmon to two lowest energy levels (transmon qubit) we get the Hamiltonian in eq.\eqref{Hamiltonian::giant::atom} up to a normalization factor of $N^{-1}_c$ in the coupling parameter. It is worth mentioning that in the presence of the transmon qubit, the frequencies $\omega_0$ and $\xi$ in $H_{\rm{photons}}$ should be obtained using the renormalized capacitance of the transmission line, $C_{\Sigma'}$.
\begin{table}
\caption{\label{tab:params}%
Experimental parameters of the circuit model taken from Ref.~\cite{ScigliuzzoPRX2022}. 
We have set the qubit–line coupling to $g_0 \sim \xi/5 = 49.8~\mathrm{MHz}$ 
for all the calculations presented in the paper.}
\begin{ruledtabular}
\begin{tabular}{lll}
Parameter & Description & Value \\
\hline
$\omega_0/2\pi$ & Cavity resonance frequency & $5.71~\mathrm{GHz}$ \\
$\xi/2\pi$      & Hopping strength            & $249~\mathrm{MHz}$ \\
$g_0/2\pi$      & Qubit cavity- array coupling strength & $50~\mathrm{MHz}$ \\
\end{tabular}
\end{ruledtabular}
\end{table}

\section{Conclusions}
\label{conclusions}
We have presented exact analytical solutions for the spontaneous emission dynamics of quantum emitters coupled to structured photonic environments, focusing in particular on giant emitters interacting with cavity–array waveguides. By employing the exact inversion of the Laplace transform, we have gone beyond the standard Weisskopf–Wigner approximation and demonstrated how the interplay of interference and non-Markovian effects leads to qualitatively new behavior.
Our results reveal the existence of distinct bound states, both inside and outside the continuum, and show that giant emitters enable their coherent superposition, giving rise to persistent oscillatory dynamics. This provides a minimal and exactly solvable model for understanding interference-induced localization and long-lived quantum coherences in structured reservoirs.
We have outlined a feasible implementation using superconducting circuit architectures, where the required cavity–array and giant atom configurations are within reach of current experiments. This suggests that the predicted phenomena, such as bound states in the continuum and their coherent interplay with conventional bound states, could be observed and exploited in near-term devices in circuit-QED. These results may be potentially extended to giant multi-emitter systems, non-Hermitian interactions, and time-dependent driving setups, offering a route toward engineered non-Markovian environments for quantum information processing and quantum simulation with giant emitters.
\begin{acknowledgments}
This work was supported by the Secretaría de Ciencia, Humanidades, Tecnología e Innovación (SECIHTI) Mexico, under grant No. CBF2023-2024-2888, and by DGAPA-PAPIIT-UNAM under grant No. IA104625. The author gratefully acknowledges valuable discussions with Sergi Terradas, Luis M. Moreno, and David Zueco on related projects. 
\end{acknowledgments}
\appendix
\section{Residue for the BIC state}
\label{appendixA}
The BIC state corresponds to a pole inside the brach cut in the interval $[-2\xi,2\xi]$. In this case the integration path in fig.\ref{fig2}(a) can be deformed in order to account for this contribution. This results in the following residue for eq.\eqref{alpha::BIC}, 
\begin{eqnarray}
    r_m = \frac{1}{1+g_0^2\left[A(u,d)+B(u,d))\right]},
\end{eqnarray}
where,
\begin{eqnarray}
    A(u,d)=du^{2-d}\frac{\left[1-\eta(u)\right]^{d}}{8\eta(u)^2}, \quad B(u,d)=\left(1+\frac{\left[1-\eta(u)\right]^{d}}{u^d}\right)\frac{u^2}{8\eta(u)}\left(1+\frac{u^2}{\eta(u)^2}\right),
\end{eqnarray}
being $\eta(u)\equiv\sqrt{1-u^2}$, and $u^{-1}=\omega_m/2\xi$, $|u|<1$.

\nocite{*}

\bibliography{refss}% Produces the bibliography via BibTeX.

\begin{thebibliography}{62}%
\makeatletter
\providecommand \@ifxundefined [1]{%
 \@ifx{#1\undefined}
}%
\providecommand \@ifnum [1]{%
 \ifnum #1\expandafter \@firstoftwo
 \else \expandafter \@secondoftwo
 \fi
}%
\providecommand \@ifx [1]{%
 \ifx #1\expandafter \@firstoftwo
 \else \expandafter \@secondoftwo
 \fi
}%
\providecommand \natexlab [1]{#1}%
\providecommand \enquote  [1]{``#1''}%
\providecommand \bibnamefont  [1]{#1}%
\providecommand \bibfnamefont [1]{#1}%
\providecommand \citenamefont [1]{#1}%
\providecommand \href@noop [0]{\@secondoftwo}%
\providecommand \href [0]{\begingroup \@sanitize@url \@href}%
\providecommand \@href[1]{\@@startlink{#1}\@@href}%
\providecommand \@@href[1]{\endgroup#1\@@endlink}%
\providecommand \@sanitize@url [0]{\catcode `\\12\catcode `\$12\catcode `\&12\catcode `\#12\catcode `\^12\catcode `\_12\catcode `\%12\relax}%
\providecommand \@@startlink[1]{}%
\providecommand \@@endlink[0]{}%
\providecommand \url  [0]{\begingroup\@sanitize@url \@url }%
\providecommand \@url [1]{\endgroup\@href {#1}{\urlprefix }}%
\providecommand \urlprefix  [0]{URL }%
\providecommand \Eprint [0]{\href }%
\providecommand \doibase [0]{https://doi.org/}%
\providecommand \selectlanguage [0]{\@gobble}%
\providecommand \bibinfo  [0]{\@secondoftwo}%
\providecommand \bibfield  [0]{\@secondoftwo}%
\providecommand \translation [1]{[#1]}%
\providecommand \BibitemOpen [0]{}%
\providecommand \bibitemStop [0]{}%
\providecommand \bibitemNoStop [0]{.\EOS\space}%
\providecommand \EOS [0]{\spacefactor3000\relax}%
\providecommand \BibitemShut  [1]{\csname bibitem#1\endcsname}%
\let\auto@bib@innerbib\@empty
%</preamble>
\bibitem [{\citenamefont {Weiss}(2012)}]{Weiss2012}%
  \BibitemOpen
  \bibfield  {author} {\bibinfo {author} {\bibfnamefont {U.}~\bibnamefont {Weiss}},\ }\href {https://doi.org/10.1142/8334} {\emph {\bibinfo {title} {Quantum Dissipative Systems}}},\ \bibinfo {edition} {4th}\ ed.\ (\bibinfo  {publisher} {WORLD SCIENTIFIC},\ \bibinfo {year} {2012})\ \Eprint {https://arxiv.org/abs/https://www.worldscientific.com/doi/pdf/10.1142/8334} {https://www.worldscientific.com/doi/pdf/10.1142/8334} \BibitemShut {NoStop}%
\bibitem [{\citenamefont {Breuer}\ and\ \citenamefont {Petruccione}(2007)}]{BreuerBook2007}%
  \BibitemOpen
  \bibfield  {author} {\bibinfo {author} {\bibfnamefont {H.-P.}\ \bibnamefont {Breuer}}\ and\ \bibinfo {author} {\bibfnamefont {F.}~\bibnamefont {Petruccione}},\ }\href {https://doi.org/10.1093/acprof:oso/9780199213900.001.0001} {\emph {\bibinfo {title} {The Theory of Open Quantum Systems}}}\ (\bibinfo  {publisher} {Oxford University Press},\ \bibinfo {year} {2007})\BibitemShut {NoStop}%
\bibitem [{\citenamefont {de~Vega}\ and\ \citenamefont {Alonso}(2017)}]{IdeVega2017}%
  \BibitemOpen
  \bibfield  {author} {\bibinfo {author} {\bibfnamefont {I.}~\bibnamefont {de~Vega}}\ and\ \bibinfo {author} {\bibfnamefont {D.}~\bibnamefont {Alonso}},\ }\bibfield  {title} {\bibinfo {title} {Dynamics of non-markovian open quantum systems},\ }\href {https://doi.org/10.1103/RevModPhys.89.015001} {\bibfield  {journal} {\bibinfo  {journal} {Rev. Mod. Phys.}\ }\textbf {\bibinfo {volume} {89}},\ \bibinfo {pages} {015001} (\bibinfo {year} {2017})}\BibitemShut {NoStop}%
\bibitem [{\citenamefont {Weisskopf}\ and\ \citenamefont {Wigner}(1930)}]{Weisskopf1930}%
  \BibitemOpen
  \bibfield  {author} {\bibinfo {author} {\bibfnamefont {V.}~\bibnamefont {Weisskopf}}\ and\ \bibinfo {author} {\bibfnamefont {E.}~\bibnamefont {Wigner}},\ }\bibfield  {title} {\bibinfo {title} {Berechnung der nat{\"u}rlichen linienbreite auf grund der diracschen lichttheorie},\ }\href {https://doi.org/10.1007/BF01336768} {\bibfield  {journal} {\bibinfo  {journal} {Zeitschrift f{\"u}r Physik}\ }\textbf {\bibinfo {volume} {63}},\ \bibinfo {pages} {54} (\bibinfo {year} {1930})}\BibitemShut {NoStop}%
\bibitem [{\citenamefont {Gonz\'alez-Tudela}\ and\ \citenamefont {Cirac}(2017)}]{GonzalezTudela&CiracPRA2017}%
  \BibitemOpen
  \bibfield  {author} {\bibinfo {author} {\bibfnamefont {A.}~\bibnamefont {Gonz\'alez-Tudela}}\ and\ \bibinfo {author} {\bibfnamefont {J.~I.}\ \bibnamefont {Cirac}},\ }\bibfield  {title} {\bibinfo {title} {Markovian and non-markovian dynamics of quantum emitters coupled to two-dimensional structured reservoirs},\ }\href {https://doi.org/10.1103/PhysRevA.96.043811} {\bibfield  {journal} {\bibinfo  {journal} {Phys. Rev. A}\ }\textbf {\bibinfo {volume} {96}},\ \bibinfo {pages} {043811} (\bibinfo {year} {2017})}\BibitemShut {NoStop}%
\bibitem [{\citenamefont {Notomi}(2010)}]{Notomi_2010}%
  \BibitemOpen
  \bibfield  {author} {\bibinfo {author} {\bibfnamefont {M.}~\bibnamefont {Notomi}},\ }\bibfield  {title} {\bibinfo {title} {Manipulating light with strongly modulated photonic crystals},\ }\href {https://doi.org/10.1088/0034-4885/73/9/096501} {\bibfield  {journal} {\bibinfo  {journal} {Reports on Progress in Physics}\ }\textbf {\bibinfo {volume} {73}},\ \bibinfo {pages} {096501} (\bibinfo {year} {2010})}\BibitemShut {NoStop}%
\bibitem [{\citenamefont {Gonz\'alez-Tudela}\ and\ \citenamefont {Porras}(2013)}]{GonzalezTudelaPorrasprl2013}%
  \BibitemOpen
  \bibfield  {author} {\bibinfo {author} {\bibfnamefont {A.}~\bibnamefont {Gonz\'alez-Tudela}}\ and\ \bibinfo {author} {\bibfnamefont {D.}~\bibnamefont {Porras}},\ }\bibfield  {title} {\bibinfo {title} {Mesoscopic entanglement induced by spontaneous emission in solid-state quantum optics},\ }\href {https://doi.org/10.1103/PhysRevLett.110.080502} {\bibfield  {journal} {\bibinfo  {journal} {Phys. Rev. Lett.}\ }\textbf {\bibinfo {volume} {110}},\ \bibinfo {pages} {080502} (\bibinfo {year} {2013})}\BibitemShut {NoStop}%
\bibitem [{\citenamefont {Masson}\ and\ \citenamefont {Asenjo-Garcia}(2020)}]{AsenjoPRR2020}%
  \BibitemOpen
  \bibfield  {author} {\bibinfo {author} {\bibfnamefont {S.~J.}\ \bibnamefont {Masson}}\ and\ \bibinfo {author} {\bibfnamefont {A.}~\bibnamefont {Asenjo-Garcia}},\ }\bibfield  {title} {\bibinfo {title} {Atomic-waveguide quantum electrodynamics},\ }\href {https://doi.org/10.1103/PhysRevResearch.2.043213} {\bibfield  {journal} {\bibinfo  {journal} {Phys. Rev. Res.}\ }\textbf {\bibinfo {volume} {2}},\ \bibinfo {pages} {043213} (\bibinfo {year} {2020})}\BibitemShut {NoStop}%
\bibitem [{\citenamefont {Scigliuzzo}\ \emph {et~al.}(2022)\citenamefont {Scigliuzzo}, \citenamefont {Calaj\`o}, \citenamefont {Ciccarello}, \citenamefont {Perez~Lozano}, \citenamefont {Bengtsson}, \citenamefont {Scarlino}, \citenamefont {Wallraff}, \citenamefont {Chang}, \citenamefont {Delsing},\ and\ \citenamefont {Gasparinetti}}]{ScigliuzzoPRX2022}%
  \BibitemOpen
  \bibfield  {author} {\bibinfo {author} {\bibfnamefont {M.}~\bibnamefont {Scigliuzzo}}, \bibinfo {author} {\bibfnamefont {G.}~\bibnamefont {Calaj\`o}}, \bibinfo {author} {\bibfnamefont {F.}~\bibnamefont {Ciccarello}}, \bibinfo {author} {\bibfnamefont {D.}~\bibnamefont {Perez~Lozano}}, \bibinfo {author} {\bibfnamefont {A.}~\bibnamefont {Bengtsson}}, \bibinfo {author} {\bibfnamefont {P.}~\bibnamefont {Scarlino}}, \bibinfo {author} {\bibfnamefont {A.}~\bibnamefont {Wallraff}}, \bibinfo {author} {\bibfnamefont {D.}~\bibnamefont {Chang}}, \bibinfo {author} {\bibfnamefont {P.}~\bibnamefont {Delsing}},\ and\ \bibinfo {author} {\bibfnamefont {S.}~\bibnamefont {Gasparinetti}},\ }\bibfield  {title} {\bibinfo {title} {Controlling atom-photon bound states in an array of josephson-junction resonators},\ }\href {https://doi.org/10.1103/PhysRevX.12.031036} {\bibfield  {journal} {\bibinfo  {journal} {Phys. Rev. X}\ }\textbf {\bibinfo {volume} {12}},\ \bibinfo {pages} {031036} (\bibinfo {year} {2022})}\BibitemShut {NoStop}%
\bibitem [{\citenamefont {Ferreira}\ \emph {et~al.}(2021)\citenamefont {Ferreira}, \citenamefont {Banker}, \citenamefont {Sipahigil}, \citenamefont {Matheny}, \citenamefont {Keller}, \citenamefont {Kim}, \citenamefont {Mirhosseini},\ and\ \citenamefont {Painter}}]{PainterPRX2021}%
  \BibitemOpen
  \bibfield  {author} {\bibinfo {author} {\bibfnamefont {V.~S.}\ \bibnamefont {Ferreira}}, \bibinfo {author} {\bibfnamefont {J.}~\bibnamefont {Banker}}, \bibinfo {author} {\bibfnamefont {A.}~\bibnamefont {Sipahigil}}, \bibinfo {author} {\bibfnamefont {M.~H.}\ \bibnamefont {Matheny}}, \bibinfo {author} {\bibfnamefont {A.~J.}\ \bibnamefont {Keller}}, \bibinfo {author} {\bibfnamefont {E.}~\bibnamefont {Kim}}, \bibinfo {author} {\bibfnamefont {M.}~\bibnamefont {Mirhosseini}},\ and\ \bibinfo {author} {\bibfnamefont {O.}~\bibnamefont {Painter}},\ }\bibfield  {title} {\bibinfo {title} {Collapse and revival of an artificial atom coupled to a structured photonic reservoir},\ }\href {https://doi.org/10.1103/PhysRevX.11.041043} {\bibfield  {journal} {\bibinfo  {journal} {Phys. Rev. X}\ }\textbf {\bibinfo {volume} {11}},\ \bibinfo {pages} {041043} (\bibinfo {year} {2021})}\BibitemShut {NoStop}%
\bibitem [{\citenamefont {S\'anchez-Burillo}\ \emph {et~al.}(2017)\citenamefont {S\'anchez-Burillo}, \citenamefont {Zueco}, \citenamefont {Mart\'{\i}n-Moreno},\ and\ \citenamefont {Garc\'{\i}a-Ripoll}}]{SanchezBurilloPRA2017}%
  \BibitemOpen
  \bibfield  {author} {\bibinfo {author} {\bibfnamefont {E.}~\bibnamefont {S\'anchez-Burillo}}, \bibinfo {author} {\bibfnamefont {D.}~\bibnamefont {Zueco}}, \bibinfo {author} {\bibfnamefont {L.}~\bibnamefont {Mart\'{\i}n-Moreno}},\ and\ \bibinfo {author} {\bibfnamefont {J.~J.}\ \bibnamefont {Garc\'{\i}a-Ripoll}},\ }\bibfield  {title} {\bibinfo {title} {Dynamical signatures of bound states in waveguide qed},\ }\href {https://doi.org/10.1103/PhysRevA.96.023831} {\bibfield  {journal} {\bibinfo  {journal} {Phys. Rev. A}\ }\textbf {\bibinfo {volume} {96}},\ \bibinfo {pages} {023831} (\bibinfo {year} {2017})}\BibitemShut {NoStop}%
\bibitem [{\citenamefont {Calaj\'o}\ \emph {et~al.}(2016)\citenamefont {Calaj\'o}, \citenamefont {Ciccarello}, \citenamefont {Chang},\ and\ \citenamefont {Rabl}}]{CalajoPRA2016}%
  \BibitemOpen
  \bibfield  {author} {\bibinfo {author} {\bibfnamefont {G.}~\bibnamefont {Calaj\'o}}, \bibinfo {author} {\bibfnamefont {F.}~\bibnamefont {Ciccarello}}, \bibinfo {author} {\bibfnamefont {D.}~\bibnamefont {Chang}},\ and\ \bibinfo {author} {\bibfnamefont {P.}~\bibnamefont {Rabl}},\ }\bibfield  {title} {\bibinfo {title} {Atom-field dressed states in slow-light waveguide qed},\ }\href {https://doi.org/10.1103/PhysRevA.93.033833} {\bibfield  {journal} {\bibinfo  {journal} {Phys. Rev. A}\ }\textbf {\bibinfo {volume} {93}},\ \bibinfo {pages} {033833} (\bibinfo {year} {2016})}\BibitemShut {NoStop}%
\bibitem [{\citenamefont {Guo}\ \emph {et~al.}(2017)\citenamefont {Guo}, \citenamefont {Grimsmo}, \citenamefont {Kockum}, \citenamefont {Pletyukhov},\ and\ \citenamefont {Johansson}}]{GuoPRA2017}%
  \BibitemOpen
  \bibfield  {author} {\bibinfo {author} {\bibfnamefont {L.}~\bibnamefont {Guo}}, \bibinfo {author} {\bibfnamefont {A.}~\bibnamefont {Grimsmo}}, \bibinfo {author} {\bibfnamefont {A.~F.}\ \bibnamefont {Kockum}}, \bibinfo {author} {\bibfnamefont {M.}~\bibnamefont {Pletyukhov}},\ and\ \bibinfo {author} {\bibfnamefont {G.}~\bibnamefont {Johansson}},\ }\bibfield  {title} {\bibinfo {title} {Giant acoustic atom: A single quantum system with a deterministic time delay},\ }\href {https://doi.org/10.1103/PhysRevA.95.053821} {\bibfield  {journal} {\bibinfo  {journal} {Phys. Rev. A}\ }\textbf {\bibinfo {volume} {95}},\ \bibinfo {pages} {053821} (\bibinfo {year} {2017})}\BibitemShut {NoStop}%
\bibitem [{\citenamefont {Kockum}\ \emph {et~al.}(2018)\citenamefont {Kockum}, \citenamefont {Johansson},\ and\ \citenamefont {Nori}}]{KockumPRL2018}%
  \BibitemOpen
  \bibfield  {author} {\bibinfo {author} {\bibfnamefont {A.~F.}\ \bibnamefont {Kockum}}, \bibinfo {author} {\bibfnamefont {G.}~\bibnamefont {Johansson}},\ and\ \bibinfo {author} {\bibfnamefont {F.}~\bibnamefont {Nori}},\ }\bibfield  {title} {\bibinfo {title} {Decoherence-free interaction between giant atoms in waveguide quantum electrodynamics},\ }\href {https://doi.org/10.1103/PhysRevLett.120.140404} {\bibfield  {journal} {\bibinfo  {journal} {Phys. Rev. Lett.}\ }\textbf {\bibinfo {volume} {120}},\ \bibinfo {pages} {140404} (\bibinfo {year} {2018})}\BibitemShut {NoStop}%
\bibitem [{\citenamefont {Carollo}\ \emph {et~al.}(2020)\citenamefont {Carollo}, \citenamefont {Cilluffo},\ and\ \citenamefont {Ciccarello}}]{CarolloPRR2020}%
  \BibitemOpen
  \bibfield  {author} {\bibinfo {author} {\bibfnamefont {A.}~\bibnamefont {Carollo}}, \bibinfo {author} {\bibfnamefont {D.}~\bibnamefont {Cilluffo}},\ and\ \bibinfo {author} {\bibfnamefont {F.}~\bibnamefont {Ciccarello}},\ }\bibfield  {title} {\bibinfo {title} {Mechanism of decoherence-free coupling between giant atoms},\ }\href {https://doi.org/10.1103/PhysRevResearch.2.043184} {\bibfield  {journal} {\bibinfo  {journal} {Phys. Rev. Res.}\ }\textbf {\bibinfo {volume} {2}},\ \bibinfo {pages} {043184} (\bibinfo {year} {2020})}\BibitemShut {NoStop}%
\bibitem [{\citenamefont {Longhi}(2020)}]{LonghiOPL2020}%
  \BibitemOpen
  \bibfield  {author} {\bibinfo {author} {\bibfnamefont {S.}~\bibnamefont {Longhi}},\ }\bibfield  {title} {\bibinfo {title} {Photonic simulation of giant atom decay},\ }\href {https://doi.org/10.1364/OL.393578} {\bibfield  {journal} {\bibinfo  {journal} {Opt. Lett.}\ }\textbf {\bibinfo {volume} {45}},\ \bibinfo {pages} {3017} (\bibinfo {year} {2020})}\BibitemShut {NoStop}%
\bibitem [{\citenamefont {Guo}\ \emph {et~al.}(2020{\natexlab{a}})\citenamefont {Guo}, \citenamefont {Wang}, \citenamefont {Purdy},\ and\ \citenamefont {Taylor}}]{TaylorPRA2020}%
  \BibitemOpen
  \bibfield  {author} {\bibinfo {author} {\bibfnamefont {S.}~\bibnamefont {Guo}}, \bibinfo {author} {\bibfnamefont {Y.}~\bibnamefont {Wang}}, \bibinfo {author} {\bibfnamefont {T.}~\bibnamefont {Purdy}},\ and\ \bibinfo {author} {\bibfnamefont {J.}~\bibnamefont {Taylor}},\ }\bibfield  {title} {\bibinfo {title} {Beyond spontaneous emission: Giant atom bounded in the continuum},\ }\href {https://doi.org/10.1103/PhysRevA.102.033706} {\bibfield  {journal} {\bibinfo  {journal} {Phys. Rev. A}\ }\textbf {\bibinfo {volume} {102}},\ \bibinfo {pages} {033706} (\bibinfo {year} {2020}{\natexlab{a}})}\BibitemShut {NoStop}%
\bibitem [{\citenamefont {Frisk~Kockum}(2021)}]{Kockumreview2021}%
  \BibitemOpen
  \bibfield  {author} {\bibinfo {author} {\bibfnamefont {A.}~\bibnamefont {Frisk~Kockum}},\ }\bibfield  {title} {\bibinfo {title} {Quantum optics with giant atoms---the first five years},\ }in\ \href@noop {} {\emph {\bibinfo {booktitle} {International Symposium on Mathematics, Quantum Theory, and Cryptography}}},\ \bibinfo {editor} {edited by\ \bibinfo {editor} {\bibfnamefont {T.}~\bibnamefont {Takagi}}, \bibinfo {editor} {\bibfnamefont {M.}~\bibnamefont {Wakayama}}, \bibinfo {editor} {\bibfnamefont {K.}~\bibnamefont {Tanaka}}, \bibinfo {editor} {\bibfnamefont {N.}~\bibnamefont {Kunihiro}}, \bibinfo {editor} {\bibfnamefont {K.}~\bibnamefont {Kimoto}},\ and\ \bibinfo {editor} {\bibfnamefont {Y.}~\bibnamefont {Ikematsu}}}\ (\bibinfo  {publisher} {Springer Singapore},\ \bibinfo {address} {Singapore},\ \bibinfo {year} {2021})\ pp.\ \bibinfo {pages} {125--146}\BibitemShut {NoStop}%
\bibitem [{\citenamefont {Wang}\ \emph {et~al.}(2021)\citenamefont {Wang}, \citenamefont {Liu}, \citenamefont {Kockum}, \citenamefont {Li},\ and\ \citenamefont {Nori}}]{WangKockumNoriPRL2021}%
  \BibitemOpen
  \bibfield  {author} {\bibinfo {author} {\bibfnamefont {X.}~\bibnamefont {Wang}}, \bibinfo {author} {\bibfnamefont {T.}~\bibnamefont {Liu}}, \bibinfo {author} {\bibfnamefont {A.~F.}\ \bibnamefont {Kockum}}, \bibinfo {author} {\bibfnamefont {H.-R.}\ \bibnamefont {Li}},\ and\ \bibinfo {author} {\bibfnamefont {F.}~\bibnamefont {Nori}},\ }\bibfield  {title} {\bibinfo {title} {Tunable chiral bound states with giant atoms},\ }\href {https://doi.org/10.1103/PhysRevLett.126.043602} {\bibfield  {journal} {\bibinfo  {journal} {Phys. Rev. Lett.}\ }\textbf {\bibinfo {volume} {126}},\ \bibinfo {pages} {043602} (\bibinfo {year} {2021})}\BibitemShut {NoStop}%
\bibitem [{\citenamefont {Feng}\ and\ \citenamefont {Jia}(2021)}]{FengPRA2021}%
  \BibitemOpen
  \bibfield  {author} {\bibinfo {author} {\bibfnamefont {S.~L.}\ \bibnamefont {Feng}}\ and\ \bibinfo {author} {\bibfnamefont {W.~Z.}\ \bibnamefont {Jia}},\ }\bibfield  {title} {\bibinfo {title} {Manipulating single-photon transport in a waveguide-qed structure containing two giant atoms},\ }\href {https://doi.org/10.1103/PhysRevA.104.063712} {\bibfield  {journal} {\bibinfo  {journal} {Phys. Rev. A}\ }\textbf {\bibinfo {volume} {104}},\ \bibinfo {pages} {063712} (\bibinfo {year} {2021})}\BibitemShut {NoStop}%
\bibitem [{\citenamefont {Du}\ and\ \citenamefont {Li}(2021)}]{DuPRA2021}%
  \BibitemOpen
  \bibfield  {author} {\bibinfo {author} {\bibfnamefont {L.}~\bibnamefont {Du}}\ and\ \bibinfo {author} {\bibfnamefont {Y.}~\bibnamefont {Li}},\ }\bibfield  {title} {\bibinfo {title} {Single-photon frequency conversion via a giant $\mathrm{\ensuremath{\Lambda}}$-type atom},\ }\href {https://doi.org/10.1103/PhysRevA.104.023712} {\bibfield  {journal} {\bibinfo  {journal} {Phys. Rev. A}\ }\textbf {\bibinfo {volume} {104}},\ \bibinfo {pages} {023712} (\bibinfo {year} {2021})}\BibitemShut {NoStop}%
\bibitem [{\citenamefont {Yu}\ \emph {et~al.}(2021)\citenamefont {Yu}, \citenamefont {Wang},\ and\ \citenamefont {Wu}}]{HongweiPRA2021}%
  \BibitemOpen
  \bibfield  {author} {\bibinfo {author} {\bibfnamefont {H.}~\bibnamefont {Yu}}, \bibinfo {author} {\bibfnamefont {Z.}~\bibnamefont {Wang}},\ and\ \bibinfo {author} {\bibfnamefont {J.-H.}\ \bibnamefont {Wu}},\ }\bibfield  {title} {\bibinfo {title} {Entanglement preparation and nonreciprocal excitation evolution in giant atoms by controllable dissipation and coupling},\ }\href {https://doi.org/10.1103/PhysRevA.104.013720} {\bibfield  {journal} {\bibinfo  {journal} {Phys. Rev. A}\ }\textbf {\bibinfo {volume} {104}},\ \bibinfo {pages} {013720} (\bibinfo {year} {2021})}\BibitemShut {NoStop}%
\bibitem [{\citenamefont {Du}\ \emph {et~al.}(2022)\citenamefont {Du}, \citenamefont {Zhang}, \citenamefont {Wu}, \citenamefont {Kockum},\ and\ \citenamefont {Li}}]{KockumPRL2022}%
  \BibitemOpen
  \bibfield  {author} {\bibinfo {author} {\bibfnamefont {L.}~\bibnamefont {Du}}, \bibinfo {author} {\bibfnamefont {Y.}~\bibnamefont {Zhang}}, \bibinfo {author} {\bibfnamefont {J.-H.}\ \bibnamefont {Wu}}, \bibinfo {author} {\bibfnamefont {A.~F.}\ \bibnamefont {Kockum}},\ and\ \bibinfo {author} {\bibfnamefont {Y.}~\bibnamefont {Li}},\ }\bibfield  {title} {\bibinfo {title} {Giant atoms in a synthetic frequency dimension},\ }\href {https://doi.org/10.1103/PhysRevLett.128.223602} {\bibfield  {journal} {\bibinfo  {journal} {Phys. Rev. Lett.}\ }\textbf {\bibinfo {volume} {128}},\ \bibinfo {pages} {223602} (\bibinfo {year} {2022})}\BibitemShut {NoStop}%
\bibitem [{\citenamefont {Soro}\ and\ \citenamefont {Kockum}(2022)}]{KockumPRA2022}%
  \BibitemOpen
  \bibfield  {author} {\bibinfo {author} {\bibfnamefont {A.}~\bibnamefont {Soro}}\ and\ \bibinfo {author} {\bibfnamefont {A.~F.}\ \bibnamefont {Kockum}},\ }\bibfield  {title} {\bibinfo {title} {Chiral quantum optics with giant atoms},\ }\href {https://doi.org/10.1103/PhysRevA.105.023712} {\bibfield  {journal} {\bibinfo  {journal} {Phys. Rev. A}\ }\textbf {\bibinfo {volume} {105}},\ \bibinfo {pages} {023712} (\bibinfo {year} {2022})}\BibitemShut {NoStop}%
\bibitem [{\citenamefont {Chen}\ \emph {et~al.}(2022)\citenamefont {Chen}, \citenamefont {Du}, \citenamefont {Guo}, \citenamefont {Wang}, \citenamefont {Zhang}, \citenamefont {Li},\ and\ \citenamefont {Wu}}]{Chen2022}%
  \BibitemOpen
  \bibfield  {author} {\bibinfo {author} {\bibfnamefont {Y.-T.}\ \bibnamefont {Chen}}, \bibinfo {author} {\bibfnamefont {L.}~\bibnamefont {Du}}, \bibinfo {author} {\bibfnamefont {L.}~\bibnamefont {Guo}}, \bibinfo {author} {\bibfnamefont {Z.}~\bibnamefont {Wang}}, \bibinfo {author} {\bibfnamefont {Y.}~\bibnamefont {Zhang}}, \bibinfo {author} {\bibfnamefont {Y.}~\bibnamefont {Li}},\ and\ \bibinfo {author} {\bibfnamefont {J.-H.}\ \bibnamefont {Wu}},\ }\bibfield  {title} {\bibinfo {title} {Nonreciprocal and chiral single-photon scattering for giant atoms},\ }\href {https://doi.org/10.1038/s42005-022-00991-3} {\bibfield  {journal} {\bibinfo  {journal} {Communications Physics}\ }\textbf {\bibinfo {volume} {5}},\ \bibinfo {pages} {215} (\bibinfo {year} {2022})}\BibitemShut {NoStop}%
\bibitem [{\citenamefont {Wang}\ and\ \citenamefont {Li}(2022)}]{Wang_2022}%
  \BibitemOpen
  \bibfield  {author} {\bibinfo {author} {\bibfnamefont {X.}~\bibnamefont {Wang}}\ and\ \bibinfo {author} {\bibfnamefont {H.-R.}\ \bibnamefont {Li}},\ }\bibfield  {title} {\bibinfo {title} {Chiral quantum network with giant atoms},\ }\href {https://doi.org/10.1088/2058-9565/ac6a04} {\bibfield  {journal} {\bibinfo  {journal} {Quantum Science and Technology}\ }\textbf {\bibinfo {volume} {7}},\ \bibinfo {pages} {035007} (\bibinfo {year} {2022})}\BibitemShut {NoStop}%
\bibitem [{\citenamefont {Noachtar}\ \emph {et~al.}(2022)\citenamefont {Noachtar}, \citenamefont {Kn\"orzer},\ and\ \citenamefont {Jonsson}}]{NoatcharPRA2022}%
  \BibitemOpen
  \bibfield  {author} {\bibinfo {author} {\bibfnamefont {D.~D.}\ \bibnamefont {Noachtar}}, \bibinfo {author} {\bibfnamefont {J.}~\bibnamefont {Kn\"orzer}},\ and\ \bibinfo {author} {\bibfnamefont {R.~H.}\ \bibnamefont {Jonsson}},\ }\bibfield  {title} {\bibinfo {title} {Nonperturbative treatment of giant atoms using chain transformations},\ }\href {https://doi.org/10.1103/PhysRevA.106.013702} {\bibfield  {journal} {\bibinfo  {journal} {Phys. Rev. A}\ }\textbf {\bibinfo {volume} {106}},\ \bibinfo {pages} {013702} (\bibinfo {year} {2022})}\BibitemShut {NoStop}%
\bibitem [{\citenamefont {Terradas-Brians\'o}\ \emph {et~al.}(2022)\citenamefont {Terradas-Brians\'o}, \citenamefont {Gonz\'alez-Guti\'errez}, \citenamefont {Nori}, \citenamefont {Mart\'{\i}n-Moreno},\ and\ \citenamefont {Zueco}}]{TerradasPRA2022}%
  \BibitemOpen
  \bibfield  {author} {\bibinfo {author} {\bibfnamefont {S.}~\bibnamefont {Terradas-Brians\'o}}, \bibinfo {author} {\bibfnamefont {C.~A.}\ \bibnamefont {Gonz\'alez-Guti\'errez}}, \bibinfo {author} {\bibfnamefont {F.}~\bibnamefont {Nori}}, \bibinfo {author} {\bibfnamefont {L.}~\bibnamefont {Mart\'{\i}n-Moreno}},\ and\ \bibinfo {author} {\bibfnamefont {D.}~\bibnamefont {Zueco}},\ }\bibfield  {title} {\bibinfo {title} {Ultrastrong waveguide qed with giant atoms},\ }\href {https://doi.org/10.1103/PhysRevA.106.063717} {\bibfield  {journal} {\bibinfo  {journal} {Phys. Rev. A}\ }\textbf {\bibinfo {volume} {106}},\ \bibinfo {pages} {063717} (\bibinfo {year} {2022})}\BibitemShut {NoStop}%
\bibitem [{\citenamefont {Xiao}\ \emph {et~al.}(2022)\citenamefont {Xiao}, \citenamefont {Wang}, \citenamefont {Li}, \citenamefont {Chen},\ and\ \citenamefont {Yuan}}]{Xiao2022}%
  \BibitemOpen
  \bibfield  {author} {\bibinfo {author} {\bibfnamefont {H.}~\bibnamefont {Xiao}}, \bibinfo {author} {\bibfnamefont {L.}~\bibnamefont {Wang}}, \bibinfo {author} {\bibfnamefont {Z.-H.}\ \bibnamefont {Li}}, \bibinfo {author} {\bibfnamefont {X.}~\bibnamefont {Chen}},\ and\ \bibinfo {author} {\bibfnamefont {L.}~\bibnamefont {Yuan}},\ }\bibfield  {title} {\bibinfo {title} {Bound state in a giant atom-modulated resonators system},\ }\href {https://doi.org/10.1038/s41534-022-00591-7} {\bibfield  {journal} {\bibinfo  {journal} {npj Quantum Information}\ }\textbf {\bibinfo {volume} {8}},\ \bibinfo {pages} {80} (\bibinfo {year} {2022})}\BibitemShut {NoStop}%
\bibitem [{\citenamefont {Cheng}\ \emph {et~al.}(2022)\citenamefont {Cheng}, \citenamefont {Wang},\ and\ \citenamefont {Liu}}]{ChengPRA2022}%
  \BibitemOpen
  \bibfield  {author} {\bibinfo {author} {\bibfnamefont {W.}~\bibnamefont {Cheng}}, \bibinfo {author} {\bibfnamefont {Z.}~\bibnamefont {Wang}},\ and\ \bibinfo {author} {\bibfnamefont {Y.-x.}\ \bibnamefont {Liu}},\ }\bibfield  {title} {\bibinfo {title} {Topology and retardation effect of a giant atom in a topological waveguide},\ }\href {https://doi.org/10.1103/PhysRevA.106.033522} {\bibfield  {journal} {\bibinfo  {journal} {Phys. Rev. A}\ }\textbf {\bibinfo {volume} {106}},\ \bibinfo {pages} {033522} (\bibinfo {year} {2022})}\BibitemShut {NoStop}%
\bibitem [{\citenamefont {Du}\ \emph {et~al.}(2023{\natexlab{a}})\citenamefont {Du}, \citenamefont {Guo},\ and\ \citenamefont {Li}}]{GuoPRA2023}%
  \BibitemOpen
  \bibfield  {author} {\bibinfo {author} {\bibfnamefont {L.}~\bibnamefont {Du}}, \bibinfo {author} {\bibfnamefont {L.}~\bibnamefont {Guo}},\ and\ \bibinfo {author} {\bibfnamefont {Y.}~\bibnamefont {Li}},\ }\bibfield  {title} {\bibinfo {title} {Complex decoherence-free interactions between giant atoms},\ }\href {https://doi.org/10.1103/PhysRevA.107.023705} {\bibfield  {journal} {\bibinfo  {journal} {Phys. Rev. A}\ }\textbf {\bibinfo {volume} {107}},\ \bibinfo {pages} {023705} (\bibinfo {year} {2023}{\natexlab{a}})}\BibitemShut {NoStop}%
\bibitem [{\citenamefont {Soro}\ \emph {et~al.}(2023)\citenamefont {Soro}, \citenamefont {Mu\~noz},\ and\ \citenamefont {Kockum}}]{SoroKockumPRA2023}%
  \BibitemOpen
  \bibfield  {author} {\bibinfo {author} {\bibfnamefont {A.}~\bibnamefont {Soro}}, \bibinfo {author} {\bibfnamefont {C.~S.}\ \bibnamefont {Mu\~noz}},\ and\ \bibinfo {author} {\bibfnamefont {A.~F.}\ \bibnamefont {Kockum}},\ }\bibfield  {title} {\bibinfo {title} {Interaction between giant atoms in a one-dimensional structured environment},\ }\href {https://doi.org/10.1103/PhysRevA.107.013710} {\bibfield  {journal} {\bibinfo  {journal} {Phys. Rev. A}\ }\textbf {\bibinfo {volume} {107}},\ \bibinfo {pages} {013710} (\bibinfo {year} {2023})}\BibitemShut {NoStop}%
\bibitem [{\citenamefont {Santos}\ and\ \citenamefont {Bachelard}(2023)}]{SantosPRL2023}%
  \BibitemOpen
  \bibfield  {author} {\bibinfo {author} {\bibfnamefont {A.~C.}\ \bibnamefont {Santos}}\ and\ \bibinfo {author} {\bibfnamefont {R.}~\bibnamefont {Bachelard}},\ }\bibfield  {title} {\bibinfo {title} {Generation of maximally entangled long-lived states with giant atoms in a waveguide},\ }\href {https://doi.org/10.1103/PhysRevLett.130.053601} {\bibfield  {journal} {\bibinfo  {journal} {Phys. Rev. Lett.}\ }\textbf {\bibinfo {volume} {130}},\ \bibinfo {pages} {053601} (\bibinfo {year} {2023})}\BibitemShut {NoStop}%
\bibitem [{\citenamefont {Yin}\ and\ \citenamefont {Liao}(2023)}]{YinPRA2023}%
  \BibitemOpen
  \bibfield  {author} {\bibinfo {author} {\bibfnamefont {X.-L.}\ \bibnamefont {Yin}}\ and\ \bibinfo {author} {\bibfnamefont {J.-Q.}\ \bibnamefont {Liao}},\ }\bibfield  {title} {\bibinfo {title} {Generation of two-giant-atom entanglement in waveguide-qed systems},\ }\href {https://doi.org/10.1103/PhysRevA.108.023728} {\bibfield  {journal} {\bibinfo  {journal} {Phys. Rev. A}\ }\textbf {\bibinfo {volume} {108}},\ \bibinfo {pages} {023728} (\bibinfo {year} {2023})}\BibitemShut {NoStop}%
\bibitem [{\citenamefont {Peng}\ and\ \citenamefont {Jia}(2023)}]{PengPRA2023}%
  \BibitemOpen
  \bibfield  {author} {\bibinfo {author} {\bibfnamefont {Y.~P.}\ \bibnamefont {Peng}}\ and\ \bibinfo {author} {\bibfnamefont {W.~Z.}\ \bibnamefont {Jia}},\ }\bibfield  {title} {\bibinfo {title} {Single-photon scattering from a chain of giant atoms coupled to a one-dimensional waveguide},\ }\href {https://doi.org/10.1103/PhysRevA.108.043709} {\bibfield  {journal} {\bibinfo  {journal} {Phys. Rev. A}\ }\textbf {\bibinfo {volume} {108}},\ \bibinfo {pages} {043709} (\bibinfo {year} {2023})}\BibitemShut {NoStop}%
\bibitem [{\citenamefont {Gu}\ \emph {et~al.}(2023)\citenamefont {Gu}, \citenamefont {Huang}, \citenamefont {Yi}, \citenamefont {Chen}, \citenamefont {Sun},\ and\ \citenamefont {Tan}}]{HuatangPRA2023}%
  \BibitemOpen
  \bibfield  {author} {\bibinfo {author} {\bibfnamefont {W.}~\bibnamefont {Gu}}, \bibinfo {author} {\bibfnamefont {H.}~\bibnamefont {Huang}}, \bibinfo {author} {\bibfnamefont {Z.}~\bibnamefont {Yi}}, \bibinfo {author} {\bibfnamefont {L.}~\bibnamefont {Chen}}, \bibinfo {author} {\bibfnamefont {L.}~\bibnamefont {Sun}},\ and\ \bibinfo {author} {\bibfnamefont {H.}~\bibnamefont {Tan}},\ }\bibfield  {title} {\bibinfo {title} {Correlated two-photon scattering in a one-dimensional waveguide coupled to two- or three-level giant atoms},\ }\href {https://doi.org/10.1103/PhysRevA.108.053718} {\bibfield  {journal} {\bibinfo  {journal} {Phys. Rev. A}\ }\textbf {\bibinfo {volume} {108}},\ \bibinfo {pages} {053718} (\bibinfo {year} {2023})}\BibitemShut {NoStop}%
\bibitem [{\citenamefont {Du}\ \emph {et~al.}(2023{\natexlab{b}})\citenamefont {Du}, \citenamefont {Chen}, \citenamefont {Zhang}, \citenamefont {Li},\ and\ \citenamefont {Wu}}]{Du_2023}%
  \BibitemOpen
  \bibfield  {author} {\bibinfo {author} {\bibfnamefont {L.}~\bibnamefont {Du}}, \bibinfo {author} {\bibfnamefont {Y.-T.}\ \bibnamefont {Chen}}, \bibinfo {author} {\bibfnamefont {Y.}~\bibnamefont {Zhang}}, \bibinfo {author} {\bibfnamefont {Y.}~\bibnamefont {Li}},\ and\ \bibinfo {author} {\bibfnamefont {J.-H.}\ \bibnamefont {Wu}},\ }\bibfield  {title} {\bibinfo {title} {Decay dynamics of a giant atom in a structured bath with broken time-reversal symmetry},\ }\href {https://doi.org/10.1088/2058-9565/ace54c} {\bibfield  {journal} {\bibinfo  {journal} {Quantum Science and Technology}\ }\textbf {\bibinfo {volume} {8}},\ \bibinfo {pages} {045010} (\bibinfo {year} {2023}{\natexlab{b}})}\BibitemShut {NoStop}%
\bibitem [{\citenamefont {Zheng}\ \emph {et~al.}(2023)\citenamefont {Zheng}, \citenamefont {Zhang}, \citenamefont {Wang}, \citenamefont {Han},\ and\ \citenamefont {Wang}}]{Zheng_2023}%
  \BibitemOpen
  \bibfield  {author} {\bibinfo {author} {\bibfnamefont {C.-M.}\ \bibnamefont {Zheng}}, \bibinfo {author} {\bibfnamefont {W.}~\bibnamefont {Zhang}}, \bibinfo {author} {\bibfnamefont {D.-Y.}\ \bibnamefont {Wang}}, \bibinfo {author} {\bibfnamefont {X.}~\bibnamefont {Han}},\ and\ \bibinfo {author} {\bibfnamefont {H.-F.}\ \bibnamefont {Wang}},\ }\bibfield  {title} {\bibinfo {title} {Simultaneously enhanced photon blockades in two microwave cavities via driving a giant atom},\ }\href {https://doi.org/10.1088/1367-2630/accd8c} {\bibfield  {journal} {\bibinfo  {journal} {New Journal of Physics}\ }\textbf {\bibinfo {volume} {25}},\ \bibinfo {pages} {043030} (\bibinfo {year} {2023})}\BibitemShut {NoStop}%
\bibitem [{\citenamefont {Zhang}\ \emph {et~al.}(2023)\citenamefont {Zhang}, \citenamefont {Liu}, \citenamefont {Gong},\ and\ \citenamefont {Wang}}]{ZhangPRA2023}%
  \BibitemOpen
  \bibfield  {author} {\bibinfo {author} {\bibfnamefont {X.}~\bibnamefont {Zhang}}, \bibinfo {author} {\bibfnamefont {C.}~\bibnamefont {Liu}}, \bibinfo {author} {\bibfnamefont {Z.}~\bibnamefont {Gong}},\ and\ \bibinfo {author} {\bibfnamefont {Z.}~\bibnamefont {Wang}},\ }\bibfield  {title} {\bibinfo {title} {Quantum interference and controllable magic cavity qed via a giant atom in a coupled resonator waveguide},\ }\href {https://doi.org/10.1103/PhysRevA.108.013704} {\bibfield  {journal} {\bibinfo  {journal} {Phys. Rev. A}\ }\textbf {\bibinfo {volume} {108}},\ \bibinfo {pages} {013704} (\bibinfo {year} {2023})}\BibitemShut {NoStop}%
\bibitem [{\citenamefont {Lim}\ \emph {et~al.}(2023)\citenamefont {Lim}, \citenamefont {Mok},\ and\ \citenamefont {Kwek}}]{Leong-Chuang2023}%
  \BibitemOpen
  \bibfield  {author} {\bibinfo {author} {\bibfnamefont {K.~H.}\ \bibnamefont {Lim}}, \bibinfo {author} {\bibfnamefont {W.-K.}\ \bibnamefont {Mok}},\ and\ \bibinfo {author} {\bibfnamefont {L.-C.}\ \bibnamefont {Kwek}},\ }\bibfield  {title} {\bibinfo {title} {Oscillating bound states in non-markovian photonic lattices},\ }\href {https://doi.org/10.1103/PhysRevA.107.023716} {\bibfield  {journal} {\bibinfo  {journal} {Phys. Rev. A}\ }\textbf {\bibinfo {volume} {107}},\ \bibinfo {pages} {023716} (\bibinfo {year} {2023})}\BibitemShut {NoStop}%
\bibitem [{\citenamefont {Roccati}\ and\ \citenamefont {Cilluffo}(2024)}]{RoccatiPRL2024}%
  \BibitemOpen
  \bibfield  {author} {\bibinfo {author} {\bibfnamefont {F.}~\bibnamefont {Roccati}}\ and\ \bibinfo {author} {\bibfnamefont {D.}~\bibnamefont {Cilluffo}},\ }\bibfield  {title} {\bibinfo {title} {Controlling markovianity with chiral giant atoms},\ }\href {https://doi.org/10.1103/PhysRevLett.133.063603} {\bibfield  {journal} {\bibinfo  {journal} {Phys. Rev. Lett.}\ }\textbf {\bibinfo {volume} {133}},\ \bibinfo {pages} {063603} (\bibinfo {year} {2024})}\BibitemShut {NoStop}%
\bibitem [{\citenamefont {Wang}\ \emph {et~al.}(2024{\natexlab{a}})\citenamefont {Wang}, \citenamefont {Zhao}, \citenamefont {Yan}, \citenamefont {Yang}, \citenamefont {Wang},\ and\ \citenamefont {Zhou}}]{WangPRA2024}%
  \BibitemOpen
  \bibfield  {author} {\bibinfo {author} {\bibfnamefont {D.-W.}\ \bibnamefont {Wang}}, \bibinfo {author} {\bibfnamefont {C.}~\bibnamefont {Zhao}}, \bibinfo {author} {\bibfnamefont {Y.-T.}\ \bibnamefont {Yan}}, \bibinfo {author} {\bibfnamefont {J.}~\bibnamefont {Yang}}, \bibinfo {author} {\bibfnamefont {Z.}~\bibnamefont {Wang}},\ and\ \bibinfo {author} {\bibfnamefont {L.}~\bibnamefont {Zhou}},\ }\bibfield  {title} {\bibinfo {title} {Topology-dependent giant-atom interaction in a topological waveguide qed system},\ }\href {https://doi.org/10.1103/PhysRevA.109.053720} {\bibfield  {journal} {\bibinfo  {journal} {Phys. Rev. A}\ }\textbf {\bibinfo {volume} {109}},\ \bibinfo {pages} {053720} (\bibinfo {year} {2024}{\natexlab{a}})}\BibitemShut {NoStop}%
\bibitem [{\citenamefont {Raaholt~Ingelsten}\ \emph {et~al.}(2024)\citenamefont {Raaholt~Ingelsten}, \citenamefont {Kockum},\ and\ \citenamefont {Soro}}]{SoroPRR2024}%
  \BibitemOpen
  \bibfield  {author} {\bibinfo {author} {\bibfnamefont {E.}~\bibnamefont {Raaholt~Ingelsten}}, \bibinfo {author} {\bibfnamefont {A.~F.}\ \bibnamefont {Kockum}},\ and\ \bibinfo {author} {\bibfnamefont {A.}~\bibnamefont {Soro}},\ }\bibfield  {title} {\bibinfo {title} {Avoiding decoherence with giant atoms in a two-dimensional structured environment},\ }\href {https://doi.org/10.1103/PhysRevResearch.6.043222} {\bibfield  {journal} {\bibinfo  {journal} {Phys. Rev. Res.}\ }\textbf {\bibinfo {volume} {6}},\ \bibinfo {pages} {043222} (\bibinfo {year} {2024})}\BibitemShut {NoStop}%
\bibitem [{\citenamefont {Wang}\ \emph {et~al.}(2024{\natexlab{b}})\citenamefont {Wang}, \citenamefont {Zhu}, \citenamefont {Liu},\ and\ \citenamefont {Nori}}]{NoriPRR224}%
  \BibitemOpen
  \bibfield  {author} {\bibinfo {author} {\bibfnamefont {X.}~\bibnamefont {Wang}}, \bibinfo {author} {\bibfnamefont {H.-B.}\ \bibnamefont {Zhu}}, \bibinfo {author} {\bibfnamefont {T.}~\bibnamefont {Liu}},\ and\ \bibinfo {author} {\bibfnamefont {F.}~\bibnamefont {Nori}},\ }\bibfield  {title} {\bibinfo {title} {Realizing quantum optics in structured environments with giant atoms},\ }\href {https://doi.org/10.1103/PhysRevResearch.6.013279} {\bibfield  {journal} {\bibinfo  {journal} {Phys. Rev. Res.}\ }\textbf {\bibinfo {volume} {6}},\ \bibinfo {pages} {013279} (\bibinfo {year} {2024}{\natexlab{b}})}\BibitemShut {NoStop}%
\bibitem [{\citenamefont {Luo}\ \emph {et~al.}(2024)\citenamefont {Luo}, \citenamefont {Yin},\ and\ \citenamefont {Liao}}]{LuoAQT2024}%
  \BibitemOpen
  \bibfield  {author} {\bibinfo {author} {\bibfnamefont {W.-B.}\ \bibnamefont {Luo}}, \bibinfo {author} {\bibfnamefont {X.-L.}\ \bibnamefont {Yin}},\ and\ \bibinfo {author} {\bibfnamefont {J.-Q.}\ \bibnamefont {Liao}},\ }\bibfield  {title} {\bibinfo {title} {Entangling two giant atoms via a topological waveguide},\ }\href {https://doi.org/https://doi.org/10.1002/qute.202400030} {\bibfield  {journal} {\bibinfo  {journal} {Advanced Quantum Technologies}\ }\textbf {\bibinfo {volume} {7}},\ \bibinfo {pages} {2400030} (\bibinfo {year} {2024})},\ \Eprint {https://arxiv.org/abs/https://advanced.onlinelibrary.wiley.com/doi/pdf/10.1002/qute.202400030} {https://advanced.onlinelibrary.wiley.com/doi/pdf/10.1002/qute.202400030} \BibitemShut {NoStop}%
\bibitem [{\citenamefont {Li}\ and\ \citenamefont {Shen}(2024)}]{LiPRA2024}%
  \BibitemOpen
  \bibfield  {author} {\bibinfo {author} {\bibfnamefont {Z.~Y.}\ \bibnamefont {Li}}\ and\ \bibinfo {author} {\bibfnamefont {H.~Z.}\ \bibnamefont {Shen}},\ }\bibfield  {title} {\bibinfo {title} {Non-markovian dynamics with a giant atom coupled to a semi-infinite photonic waveguide},\ }\href {https://doi.org/10.1103/PhysRevA.109.023712} {\bibfield  {journal} {\bibinfo  {journal} {Phys. Rev. A}\ }\textbf {\bibinfo {volume} {109}},\ \bibinfo {pages} {023712} (\bibinfo {year} {2024})}\BibitemShut {NoStop}%
\bibitem [{\citenamefont {Gu}\ \emph {et~al.}(2024{\natexlab{a}})\citenamefont {Gu}, \citenamefont {Chen}, \citenamefont {Yi}, \citenamefont {Liu},\ and\ \citenamefont {Li}}]{GaoxiangPRA2024}%
  \BibitemOpen
  \bibfield  {author} {\bibinfo {author} {\bibfnamefont {W.}~\bibnamefont {Gu}}, \bibinfo {author} {\bibfnamefont {L.}~\bibnamefont {Chen}}, \bibinfo {author} {\bibfnamefont {Z.}~\bibnamefont {Yi}}, \bibinfo {author} {\bibfnamefont {S.}~\bibnamefont {Liu}},\ and\ \bibinfo {author} {\bibfnamefont {G.-x.}\ \bibnamefont {Li}},\ }\bibfield  {title} {\bibinfo {title} {Tunable photon-photon correlations in waveguide qed systems with giant atoms},\ }\href {https://doi.org/10.1103/PhysRevA.109.023720} {\bibfield  {journal} {\bibinfo  {journal} {Phys. Rev. A}\ }\textbf {\bibinfo {volume} {109}},\ \bibinfo {pages} {023720} (\bibinfo {year} {2024}{\natexlab{a}})}\BibitemShut {NoStop}%
\bibitem [{\citenamefont {Xu}\ and\ \citenamefont {Guo}(2024)}]{Xu_2024}%
  \BibitemOpen
  \bibfield  {author} {\bibinfo {author} {\bibfnamefont {L.}~\bibnamefont {Xu}}\ and\ \bibinfo {author} {\bibfnamefont {L.}~\bibnamefont {Guo}},\ }\bibfield  {title} {\bibinfo {title} {Catch and release of propagating bosonic field with non-markovian giant atom},\ }\href {https://doi.org/10.1088/1367-2630/ad18ed} {\bibfield  {journal} {\bibinfo  {journal} {New Journal of Physics}\ }\textbf {\bibinfo {volume} {26}},\ \bibinfo {pages} {013025} (\bibinfo {year} {2024})}\BibitemShut {NoStop}%
\bibitem [{\citenamefont {Li}\ \emph {et~al.}(2024)\citenamefont {Li}, \citenamefont {Zhang}, \citenamefont {Du}, \citenamefont {Li},\ and\ \citenamefont {Wu}}]{HuaizhiPRA2024}%
  \BibitemOpen
  \bibfield  {author} {\bibinfo {author} {\bibfnamefont {S.-Y.}\ \bibnamefont {Li}}, \bibinfo {author} {\bibfnamefont {Z.-Q.}\ \bibnamefont {Zhang}}, \bibinfo {author} {\bibfnamefont {L.}~\bibnamefont {Du}}, \bibinfo {author} {\bibfnamefont {Y.}~\bibnamefont {Li}},\ and\ \bibinfo {author} {\bibfnamefont {H.}~\bibnamefont {Wu}},\ }\bibfield  {title} {\bibinfo {title} {Single-photon scattering in giant-atom waveguide systems with chiral coupling},\ }\href {https://doi.org/10.1103/PhysRevA.109.063703} {\bibfield  {journal} {\bibinfo  {journal} {Phys. Rev. A}\ }\textbf {\bibinfo {volume} {109}},\ \bibinfo {pages} {063703} (\bibinfo {year} {2024})}\BibitemShut {NoStop}%
\bibitem [{\citenamefont {Gu}\ \emph {et~al.}(2024{\natexlab{b}})\citenamefont {Gu}, \citenamefont {Li}, \citenamefont {Tian}, \citenamefont {Yi},\ and\ \citenamefont {Li}}]{GaoXiang2PRA2024}%
  \BibitemOpen
  \bibfield  {author} {\bibinfo {author} {\bibfnamefont {W.}~\bibnamefont {Gu}}, \bibinfo {author} {\bibfnamefont {T.}~\bibnamefont {Li}}, \bibinfo {author} {\bibfnamefont {Y.}~\bibnamefont {Tian}}, \bibinfo {author} {\bibfnamefont {Z.}~\bibnamefont {Yi}},\ and\ \bibinfo {author} {\bibfnamefont {G.-x.}\ \bibnamefont {Li}},\ }\bibfield  {title} {\bibinfo {title} {Two-photon dynamics in non-markovian waveguide qed with a giant atom},\ }\href {https://doi.org/10.1103/PhysRevA.110.033707} {\bibfield  {journal} {\bibinfo  {journal} {Phys. Rev. A}\ }\textbf {\bibinfo {volume} {110}},\ \bibinfo {pages} {033707} (\bibinfo {year} {2024}{\natexlab{b}})}\BibitemShut {NoStop}%
\bibitem [{\citenamefont {Weng}\ \emph {et~al.}(2024)\citenamefont {Weng}, \citenamefont {Wang},\ and\ \citenamefont {Wang}}]{WengPRA2024}%
  \BibitemOpen
  \bibfield  {author} {\bibinfo {author} {\bibfnamefont {M.}~\bibnamefont {Weng}}, \bibinfo {author} {\bibfnamefont {X.}~\bibnamefont {Wang}},\ and\ \bibinfo {author} {\bibfnamefont {Z.}~\bibnamefont {Wang}},\ }\bibfield  {title} {\bibinfo {title} {Interaction and entanglement engineering in a driven-giant-atom setup with a coupled resonator waveguide},\ }\href {https://doi.org/10.1103/PhysRevA.110.023721} {\bibfield  {journal} {\bibinfo  {journal} {Phys. Rev. A}\ }\textbf {\bibinfo {volume} {110}},\ \bibinfo {pages} {023721} (\bibinfo {year} {2024})}\BibitemShut {NoStop}%
\bibitem [{\citenamefont {Guo}\ \emph {et~al.}(2024)\citenamefont {Guo}, \citenamefont {Zhu}, \citenamefont {Guo}, \citenamefont {Lin}, \citenamefont {Li},\ and\ \citenamefont {Tu}}]{TaoPRA2024}%
  \BibitemOpen
  \bibfield  {author} {\bibinfo {author} {\bibfnamefont {A.-L.}\ \bibnamefont {Guo}}, \bibinfo {author} {\bibfnamefont {L.-T.}\ \bibnamefont {Zhu}}, \bibinfo {author} {\bibfnamefont {G.-C.}\ \bibnamefont {Guo}}, \bibinfo {author} {\bibfnamefont {Z.-R.}\ \bibnamefont {Lin}}, \bibinfo {author} {\bibfnamefont {C.-F.}\ \bibnamefont {Li}},\ and\ \bibinfo {author} {\bibfnamefont {T.}~\bibnamefont {Tu}},\ }\bibfield  {title} {\bibinfo {title} {Phonon superradiance with time delays from collective giant atoms},\ }\href {https://doi.org/10.1103/PhysRevA.109.033711} {\bibfield  {journal} {\bibinfo  {journal} {Phys. Rev. A}\ }\textbf {\bibinfo {volume} {109}},\ \bibinfo {pages} {033711} (\bibinfo {year} {2024})}\BibitemShut {NoStop}%
\bibitem [{\citenamefont {Gao}\ \emph {et~al.}(2024)\citenamefont {Gao}, \citenamefont {Li}, \citenamefont {Li}, \citenamefont {Liu},\ and\ \citenamefont {Wang}}]{XinPRA2024}%
  \BibitemOpen
  \bibfield  {author} {\bibinfo {author} {\bibfnamefont {Z.-M.}\ \bibnamefont {Gao}}, \bibinfo {author} {\bibfnamefont {J.-Q.}\ \bibnamefont {Li}}, \bibinfo {author} {\bibfnamefont {Z.-W.}\ \bibnamefont {Li}}, \bibinfo {author} {\bibfnamefont {W.-X.}\ \bibnamefont {Liu}},\ and\ \bibinfo {author} {\bibfnamefont {X.}~\bibnamefont {Wang}},\ }\bibfield  {title} {\bibinfo {title} {Circuit qed with giant atoms coupling to left-handed superlattice metamaterials},\ }\href {https://doi.org/10.1103/PhysRevA.109.013716} {\bibfield  {journal} {\bibinfo  {journal} {Phys. Rev. A}\ }\textbf {\bibinfo {volume} {109}},\ \bibinfo {pages} {013716} (\bibinfo {year} {2024})}\BibitemShut {NoStop}%
\bibitem [{\citenamefont {Leonforte}\ \emph {et~al.}(2024)\citenamefont {Leonforte}, \citenamefont {Sun}, \citenamefont {Valenti}, \citenamefont {Spagnolo}, \citenamefont {Illuminati}, \citenamefont {Carollo},\ and\ \citenamefont {Ciccarello}}]{Leonforte_2025}%
  \BibitemOpen
  \bibfield  {author} {\bibinfo {author} {\bibfnamefont {L.}~\bibnamefont {Leonforte}}, \bibinfo {author} {\bibfnamefont {X.}~\bibnamefont {Sun}}, \bibinfo {author} {\bibfnamefont {D.}~\bibnamefont {Valenti}}, \bibinfo {author} {\bibfnamefont {B.}~\bibnamefont {Spagnolo}}, \bibinfo {author} {\bibfnamefont {F.}~\bibnamefont {Illuminati}}, \bibinfo {author} {\bibfnamefont {A.}~\bibnamefont {Carollo}},\ and\ \bibinfo {author} {\bibfnamefont {F.}~\bibnamefont {Ciccarello}},\ }\bibfield  {title} {\bibinfo {title} {Quantum optics with giant atoms in a structured photonic bath},\ }\href {https://doi.org/10.1088/2058-9565/ada08d} {\bibfield  {journal} {\bibinfo  {journal} {Quantum Science and Technology}\ }\textbf {\bibinfo {volume} {10}},\ \bibinfo {pages} {015057} (\bibinfo {year} {2024})}\BibitemShut {NoStop}%
\bibitem [{\citenamefont {Chen}\ and\ \citenamefont {Frisk~Kockum}(2025)}]{Chen_2025}%
  \BibitemOpen
  \bibfield  {author} {\bibinfo {author} {\bibfnamefont {G.}~\bibnamefont {Chen}}\ and\ \bibinfo {author} {\bibfnamefont {A.}~\bibnamefont {Frisk~Kockum}},\ }\bibfield  {title} {\bibinfo {title} {Simulating open quantum systems with giant atoms},\ }\href {https://doi.org/10.1088/2058-9565/adb2bd} {\bibfield  {journal} {\bibinfo  {journal} {Quantum Science and Technology}\ }\textbf {\bibinfo {volume} {10}},\ \bibinfo {pages} {025028} (\bibinfo {year} {2025})}\BibitemShut {NoStop}%
\bibitem [{\citenamefont {Kannan}\ \emph {et~al.}(2020)\citenamefont {Kannan}, \citenamefont {Ruckriegel}, \citenamefont {Campbell}, \citenamefont {Frisk~Kockum}, \citenamefont {Braum{\"u}ller}, \citenamefont {Kim}, \citenamefont {Kjaergaard}, \citenamefont {Krantz}, \citenamefont {Melville}, \citenamefont {Niedzielski}, \citenamefont {Veps{\"a}l{\"a}inen}, \citenamefont {Winik}, \citenamefont {Yoder}, \citenamefont {Nori}, \citenamefont {Orlando}, \citenamefont {Gustavsson},\ and\ \citenamefont {Oliver}}]{Kannan2020}%
  \BibitemOpen
  \bibfield  {author} {\bibinfo {author} {\bibfnamefont {B.}~\bibnamefont {Kannan}}, \bibinfo {author} {\bibfnamefont {M.~J.}\ \bibnamefont {Ruckriegel}}, \bibinfo {author} {\bibfnamefont {D.~L.}\ \bibnamefont {Campbell}}, \bibinfo {author} {\bibfnamefont {A.}~\bibnamefont {Frisk~Kockum}}, \bibinfo {author} {\bibfnamefont {J.}~\bibnamefont {Braum{\"u}ller}}, \bibinfo {author} {\bibfnamefont {D.~K.}\ \bibnamefont {Kim}}, \bibinfo {author} {\bibfnamefont {M.}~\bibnamefont {Kjaergaard}}, \bibinfo {author} {\bibfnamefont {P.}~\bibnamefont {Krantz}}, \bibinfo {author} {\bibfnamefont {A.}~\bibnamefont {Melville}}, \bibinfo {author} {\bibfnamefont {B.~M.}\ \bibnamefont {Niedzielski}}, \bibinfo {author} {\bibfnamefont {A.}~\bibnamefont {Veps{\"a}l{\"a}inen}}, \bibinfo {author} {\bibfnamefont {R.}~\bibnamefont {Winik}}, \bibinfo {author} {\bibfnamefont {J.~L.}\ \bibnamefont {Yoder}}, \bibinfo {author} {\bibfnamefont {F.}~\bibnamefont {Nori}}, \bibinfo {author} {\bibfnamefont {T.~P.}\ \bibnamefont {Orlando}}, \bibinfo
  {author} {\bibfnamefont {S.}~\bibnamefont {Gustavsson}},\ and\ \bibinfo {author} {\bibfnamefont {W.~D.}\ \bibnamefont {Oliver}},\ }\bibfield  {title} {\bibinfo {title} {Waveguide quantum electrodynamics with superconducting artificial giant atoms},\ }\href {https://doi.org/10.1038/s41586-020-2529-9} {\bibfield  {journal} {\bibinfo  {journal} {Nature}\ }\textbf {\bibinfo {volume} {583}},\ \bibinfo {pages} {775} (\bibinfo {year} {2020})}\BibitemShut {NoStop}%
\bibitem [{\citenamefont {Azzam}\ and\ \citenamefont {Kildishev}(2021)}]{AzzamAOM2021}%
  \BibitemOpen
  \bibfield  {author} {\bibinfo {author} {\bibfnamefont {S.~I.}\ \bibnamefont {Azzam}}\ and\ \bibinfo {author} {\bibfnamefont {A.~V.}\ \bibnamefont {Kildishev}},\ }\bibfield  {title} {\bibinfo {title} {Photonic bound states in the continuum: From basics to applications},\ }\href {https://doi.org/https://doi.org/10.1002/adom.202001469} {\bibfield  {journal} {\bibinfo  {journal} {Advanced Optical Materials}\ }\textbf {\bibinfo {volume} {9}},\ \bibinfo {pages} {2001469} (\bibinfo {year} {2021})},\ \Eprint {https://arxiv.org/abs/https://advanced.onlinelibrary.wiley.com/doi/pdf/10.1002/adom.202001469} {https://advanced.onlinelibrary.wiley.com/doi/pdf/10.1002/adom.202001469} \BibitemShut {NoStop}%
\bibitem [{\citenamefont {Guo}\ \emph {et~al.}(2020{\natexlab{b}})\citenamefont {Guo}, \citenamefont {Kockum}, \citenamefont {Marquardt},\ and\ \citenamefont {Johansson}}]{KockumPRR2020}%
  \BibitemOpen
  \bibfield  {author} {\bibinfo {author} {\bibfnamefont {L.}~\bibnamefont {Guo}}, \bibinfo {author} {\bibfnamefont {A.~F.}\ \bibnamefont {Kockum}}, \bibinfo {author} {\bibfnamefont {F.}~\bibnamefont {Marquardt}},\ and\ \bibinfo {author} {\bibfnamefont {G.}~\bibnamefont {Johansson}},\ }\bibfield  {title} {\bibinfo {title} {Oscillating bound states for a giant atom},\ }\href {https://doi.org/10.1103/PhysRevResearch.2.043014} {\bibfield  {journal} {\bibinfo  {journal} {Phys. Rev. Res.}\ }\textbf {\bibinfo {volume} {2}},\ \bibinfo {pages} {043014} (\bibinfo {year} {2020}{\natexlab{b}})}\BibitemShut {NoStop}%
\bibitem [{\citenamefont {Berman}\ and\ \citenamefont {Ford}(2010)}]{BERMAN2010175}%
  \BibitemOpen
  \bibfield  {author} {\bibinfo {author} {\bibfnamefont {P.~R.}\ \bibnamefont {Berman}}\ and\ \bibinfo {author} {\bibfnamefont {G.~W.}\ \bibnamefont {Ford}},\ }\bibfield  {title} {\bibinfo {title} {Chapter 5 - spontaneous decay, unitarity, and the weisskopf–wigner approximation},\ }in\ \href {https://doi.org/https://doi.org/10.1016/S1049-250X(10)59005-0} {\emph {\bibinfo {booktitle} {Advances in Atomic, Molecular, and Optical Physics}}},\ \bibinfo {series} {Advances In Atomic, Molecular, and Optical Physics}, Vol.~\bibinfo {volume} {59},\ \bibinfo {editor} {edited by\ \bibinfo {editor} {\bibfnamefont {E.}~\bibnamefont {Arimondo}}, \bibinfo {editor} {\bibfnamefont {P.}~\bibnamefont {Berman}},\ and\ \bibinfo {editor} {\bibfnamefont {C.}~\bibnamefont {Lin}}}\ (\bibinfo  {publisher} {Academic Press},\ \bibinfo {year} {2010})\ pp.\ \bibinfo {pages} {175--221}\BibitemShut {NoStop}%
\bibitem [{\citenamefont {Longhi}(2007)}]{LonghiEPJB2007}%
  \BibitemOpen
  \bibfield  {author} {\bibinfo {author} {\bibfnamefont {S.}~\bibnamefont {Longhi}},\ }\bibfield  {title} {\bibinfo {title} {Bound states in the continuum in a single-level fano-anderson model},\ }\href {https://doi.org/10.1140/epjb/e2007-00143-2} {\bibfield  {journal} {\bibinfo  {journal} {The European Physical Journal B}\ }\textbf {\bibinfo {volume} {57}},\ \bibinfo {pages} {45} (\bibinfo {year} {2007})}\BibitemShut {NoStop}%
\bibitem [{\citenamefont {Sinha}\ \emph {et~al.}(2020)\citenamefont {Sinha}, \citenamefont {Meystre}, \citenamefont {Goldschmidt}, \citenamefont {Fatemi}, \citenamefont {Rolston},\ and\ \citenamefont {Solano}}]{SinhaPRL2020}%
  \BibitemOpen
  \bibfield  {author} {\bibinfo {author} {\bibfnamefont {K.}~\bibnamefont {Sinha}}, \bibinfo {author} {\bibfnamefont {P.}~\bibnamefont {Meystre}}, \bibinfo {author} {\bibfnamefont {E.~A.}\ \bibnamefont {Goldschmidt}}, \bibinfo {author} {\bibfnamefont {F.~K.}\ \bibnamefont {Fatemi}}, \bibinfo {author} {\bibfnamefont {S.~L.}\ \bibnamefont {Rolston}},\ and\ \bibinfo {author} {\bibfnamefont {P.}~\bibnamefont {Solano}},\ }\bibfield  {title} {\bibinfo {title} {Non-markovian collective emission from macroscopically separated emitters},\ }\href {https://doi.org/10.1103/PhysRevLett.124.043603} {\bibfield  {journal} {\bibinfo  {journal} {Phys. Rev. Lett.}\ }\textbf {\bibinfo {volume} {124}},\ \bibinfo {pages} {043603} (\bibinfo {year} {2020})}\BibitemShut {NoStop}%
\bibitem [{\citenamefont {Hall}\ \emph {et~al.}(2014)\citenamefont {Hall}, \citenamefont {Cresser}, \citenamefont {Li},\ and\ \citenamefont {Andersson}}]{AnderssonPRA2014}%
  \BibitemOpen
  \bibfield  {author} {\bibinfo {author} {\bibfnamefont {M.~J.~W.}\ \bibnamefont {Hall}}, \bibinfo {author} {\bibfnamefont {J.~D.}\ \bibnamefont {Cresser}}, \bibinfo {author} {\bibfnamefont {L.}~\bibnamefont {Li}},\ and\ \bibinfo {author} {\bibfnamefont {E.}~\bibnamefont {Andersson}},\ }\bibfield  {title} {\bibinfo {title} {Canonical form of master equations and characterization of non-markovianity},\ }\href {https://doi.org/10.1103/PhysRevA.89.042120} {\bibfield  {journal} {\bibinfo  {journal} {Phys. Rev. A}\ }\textbf {\bibinfo {volume} {89}},\ \bibinfo {pages} {042120} (\bibinfo {year} {2014})}\BibitemShut {NoStop}%
\end{thebibliography}%

\end{document}